\documentclass[
  reprint,
  prd,
  showpacs,
  amsmath,
  amssymb,
  superscriptaddress,
  longbibliography,
]{revtex4-2}

\usepackage{ragged2e}
\usepackage[usenames,dvipsnames]{color}
\usepackage{xcolor}
\usepackage{bm}
\usepackage{units}
\usepackage{graphicx}
\usepackage{braket}
\usepackage[normalem]{ulem}
\usepackage{empheq}

\newcommand{\br}{{\bf r}}

\renewcommand{\vec}[1]{\mathbf{#1}}

\usepackage{pdfpages}
\usepackage{pgffor}
\makeatletter
\AtBeginDocument{\let\LS@rot\@undefined}
\makeatother

\begin{document}

\title{Magic-angle semimetals}

\author{Yixing Fu }
\thanks{YXF, EJK, and JHW contributed equally.}
\affiliation{Department of Physics and Astronomy, Center for Materials Theory, Rutgers University, Piscataway, NJ 08854 USA}
\author{Elio J. K\"onig }
\thanks{YXF, EJK, and JHW contributed equally.}
\affiliation{Department of Physics and Astronomy, Center for Materials Theory, Rutgers University, Piscataway, NJ 08854 USA}
\author{Justin H. Wilson}
\thanks{YXF, EJK, and JHW contributed equally.}
\affiliation{Department of Physics and Astronomy, Center for Materials Theory, Rutgers University, Piscataway, NJ 08854 USA}
\affiliation{Institute of Quantum Information and Matter and Department of Physics,
California Institute of Technology, Pasadena, California 91125 USA}
\author{Yang-Zhi Chou}
\affiliation{Condensed Matter Theory Center and the Joint Quantum Institute, Department of Physics,
	University of Maryland, College Park, MD 20742 USA}
\affiliation{Department of Physics and Center for Theory of Quantum Matter, University of Colorado Boulder, Boulder, Colorado 80309, USA}
\author{Jedediah H. Pixley}
\email{Corresponding author: {\tt jed.pixley@physics.rutgers.edu}}
\affiliation{Department of Physics and Astronomy, Center for Materials Theory, Rutgers University, Piscataway, NJ 08854 USA}

\begin{abstract}
 Breakthroughs in two-dimensional van der Waals heterostructures have revealed that twisting creates a moir\'e pattern that quenches the kinetic energy of electrons, allowing for exotic many-body states.
 We show that cold-atomic, trapped ion, and metamaterial systems can emulate the effects of a twist in many models from one to three dimensions. Further, we demonstrate at larger angles (and argue at smaller angles) that by considering incommensurate effects, the magic-angle effect becomes a single-particle quantum phase transition (including in a model for twisted bilayer graphene in the chiral limit).  We call these models ``magic-angle semimetals.'' Each contains nodes in the band structure and an incommensurate modulation.  At magic-angle criticality, we report a nonanalytic density of states, flat bands, multifractal wave functions that Anderson delocalize in momentum space, and an essentially divergent effective interaction scale. As a particular example, we discuss how to observe this effect in an ultracold Fermi gas.
\end{abstract}

\maketitle

\section{Introduction.}

The engineering of band structures with non-trivial topological wave functions has achieved success in creating and controlling quantum phases in a variety of systems such as doped  semiconductors~\cite{XuHasan2011,LindnerGalitski2011,BurkovBalents2011,BelopolskiHasan2017},
ultracold atoms~\cite{DalibardOeberg2011,AidelsburgerGoldmann2017}, and metamaterials~\cite{LeeThomale2017,OzawaCarusotto2018}.
With the recent advance in twisted graphene heterostructores \cite{CaoJarillo2018a,ChenWang2018,CaoJarillo2018b,Yankowitz-2018}
(i.e. ``twistronics'') new, 
strongly interacting, solid state systems can now also be engineered with a rather weakly correlated two-dimensional semimetal (graphene) \cite{SongLevitov2013,KimTutuc2017,WuMacDonald2018}.
In these systems, as a consequence of the quenched kinetic energy, correlations dominate the physics and exotic
many-body states may form.
This interpretation relies on the reduction of the electronic velocity and large increase of the density of states (DOS) which was shown in twisted bilayer graphene (TBG) theoretically \cite{TramblyMagaud2010,BistritzerMacDonald2011,DosSantosNeto2012,SanJose2012} and experimentally \cite{LiAndrei2010,Brihuega2012,KimSmet2016} prior to the recent groundbreaking discoveries in Refs.~\cite{CaoJarillo2018a,CaoJarillo2018b,ChenWang2018}.
Understanding the essential single-particle ingredients necessary to build emulators of TBG can help shed light
on the strong coupling regime where consensus about the form of an effective low-energy description remains elusive~\cite{PadhiPhillips2018, PoSenthil2018,YuanFu2018,KangVafek2018,XuBalents2018,LianBernevig2018,DodaroWang2018}.

In this manuscript, we develop the theory for twistronic emulators
by first distilling the basic physical phenomena that create correlated flat bands out of two-dimensional Dirac cones.
Generically, \emph{quasiperiodicity} that respects the symmetry protecting the Dirac nodes creates flat bands in nodal, semimetallic band structures in a universal fashion near a previously unnoticed single particle quantum phase transition (QPT)---what we call the ``magic-angle'' in analogy to TBG.
At small angles in TBG, a single scattering wavevector accounts for the 
majority of the 
band flattening \cite{BistritzerMacDonald2011,Tarnopolsky2019} but misses any QPT.
With quasiperiodicity, an infinite sequence of higher wavevectors (i.e., Brillouin zone downfoldings) further flatten the bands and culminate into a QPT.
{ This band flattening occurs irrespective of the topology present, and in fact, many of the models we study have topology distinct from TBG~\cite{PoSenthil2018}.}
We demonstrate strong correlations by computing Wannier states within this series of bands; these lead to a Hubbard model with a quenched kinetic energy  and relative to this, the interaction scale is increased dramatically.
We therefore argue that the single particle quantum critical state is unstable towards the inclusion of interactions, which form a correlated insulator at half filling.

Crucially, our findings are independent of many of the system's details and, therefore, demonstrate the existence of a wide multitude of engineered, strongly-coupled quantum systems that we call \emph{magic-angle semimetals}.
To demonstrate this, we classify the family of these models with symmetry protected nodes (including chiral TBG at moderate twist angles)  as well as introduce and solve a series of models; most of which can be straightforwardly realized with existing ultracold atom, trapped ion, and metamaterial experimental setups.
Thus, we propose a simple route to emulate the phenomena of magic-angle TBG in a wide variety of quantum many body systems~\cite{WolfBlatter2018,WuMacDonald2018}.
Last, we show that the magic-angle effect can be observed at experimentally relevant time scales and temperatures in interacting ultracold Fermi gases through measurements of wavepacket dynamics.

\begin{figure}
\includegraphics[width=.45\textwidth]{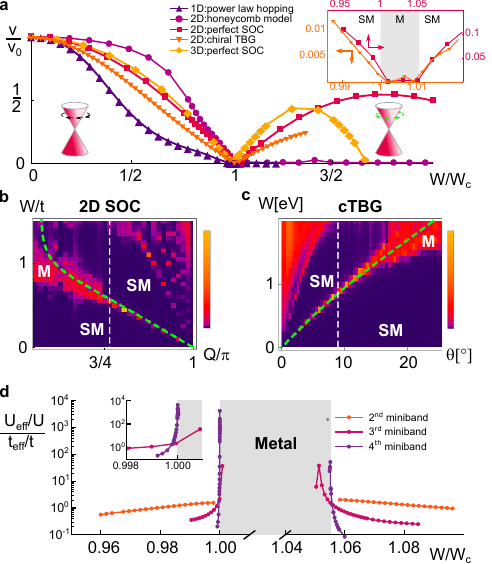}
\caption{{\bf Magic-angle transition.} A quasiperiodic potential or tunneling generically drives an eigenstate quantum phase transition from a semimetal (SM) to metal (M).
{\bf a}. For many models, the velocity at the Dirac node $v$ decreases with the strength of the potential $W$ until it reaches $v=0$ at the transition, $W_c$; this is an indication of the flattening of the bands.
In some cases an intermediate metallic phase (see inset) separates a reentrant semimetal with a reversed helicity (depicted by the Dirac cones).
{\bf b, \bf c} We construct a phase diagram in terms of potential strength $W$ (interlayer tunneling for cTBG) and quasiperiodic modulation $Q$ (twist angle $\theta$ for cTBG) by computing the density of states at zero energy $\rho(0)$; analytical perturbative results [see Eq.~\eqref{eq:vpert}, 
{Supplementary Notes 3}
and ~\cite{BistritzerMacDonald2011,Tarnopolsky2019}]
are represented by the green dashed lines.
Cuts along the dashed white lines are presented in Fig.~\ref{fig:DOSandDISP}{\bf c},{\bf d}. Color bars represent $\rho(0)$ and with widths {\bf b:} 5, and {\bf c:} 1.25 and dark purple represents the value 0 on both.
{\bf d}. An infinite number of semimetal minibands form as the transition is approached; each has higher effective interaction than the last as we approach the transition.
For 2D SOC, we construct exponentially localized Wannier states on the first four minibands (see Fig.~\ref{fig:Spectrum})
leading to a model with an effective, strongly renormalized Hubbard interaction $U_{\mathrm{eff}}/t_{\mathrm{eff}}$ in terms of the bare interaction $U/t$.
}
\label{fig:Schematics}
\end{figure}

\section{Results}

\subsection{{`Magic-angle semimetals'}.}
The whole class of magic-angle semimetal models are governed by Hamiltonians of the form
\begin{equation}
\hat H =  \hat T + \hat V +  \hat U
\label{eqn:genH}
\end{equation}
containing single particle hopping $\hat T$, a quasiperiodic modulation $\hat V$ (such as potential scattering {or interlayer tunneling}), and inter-particle interactions $\hat U$.
The kinetic term $\hat T$ has isolated nodal points in the Brillouin zone where the DOS vanishes in a power-law fashion (i.e. semimetallic).
The quasiperiodicity in $\hat V$ is encoded in an \emph{angle} originating either from  twisted bilayers or the projective construction of quasicrystals \cite{Janot2012}, and it is characterized by an amplitude $W$ and an
incommensurate modulation $Q$ {(or twist angle $\theta$)}.

Generalizing the physics of the first magic angle of TBG to magic-angle semimetals results in the phenomena summarized by Fig.~\ref{fig:Schematics}.
First, increasing $W$ quenches the kinetic energy, reducing the Dirac velocity $v$ until it ultimately reaches zero at the single-particle quantum critical point (where the DOS becomes nonanalytic).
The velocity vanishes in a universal manner characterized by critical exponents that are distinct in each dimension.
Second, the DOS and wave functions display a transition from a ballistic semimetal to a metallic phase; this is a { so-called  
`unfreezing'} transition in momentum space, which represents a {non-standard} form of delocalization~\cite{EversMirlin2008}.
For a subset of magic-angle semimetals [including Eqs.~\eqref{eq:H0} and~\eqref{eq:tbg} below], the semimetal reenters at a second transition $W_c'$ with a reversed sign of the helicity at each Dirac node~\cite{PixleyGopalakrishnan2018}; for general $Q$ (or $\theta$), multiple semimetal-metal-semimetal transitions can appear as $W$ is tuned, see Figs.~\ref{fig:Schematics}{\bf b}, {\bf c}.
Third, the quenched kinetic energy implies a divergence of the dimensionless interaction coupling constant, Fig.~\ref{fig:Schematics}{\bf d}, leading to exotic many-body states.
Importantly, these effects occur generically under the necessary condition that the quasiperiodic modulation respects the symmetries which protect the semimetallic touching points (see {Supplementary Note 3}).

\begin{figure*}[t]
\includegraphics[width=1\textwidth]{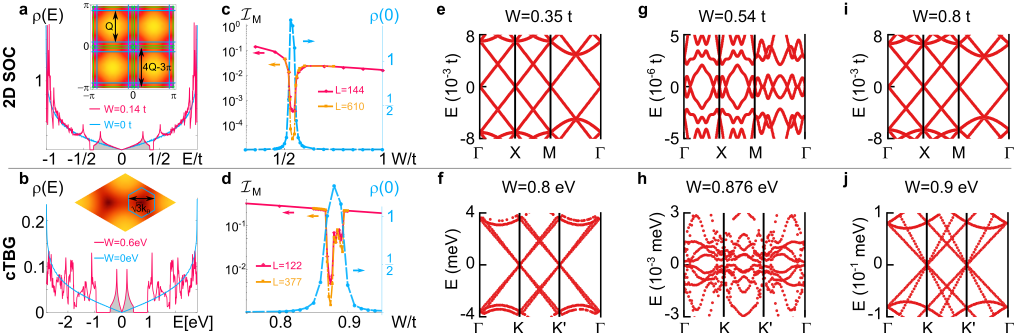}
\caption{{\bf Eigenstate transition as manifested in the single particle spectrum}. Panels {\bf a,b}: DOS $\rho(E)$ in units of $(t L^2)^{-1}$ averaged over 300 realizations of phases $\phi_\mu$  and random twisted boundary conditions (explained in more detail in {Supplementary Note 2}).
 The gray
shading represents the number of states in the first 
miniband and matches the area of the
 mini Brillouin zones around each Dirac point produced by the leading 
scattering vectors depicted in the inset of {\bf a, \bf b} (we chose a rhombic representation of the Brillouin zone of TBG such that $\mathbf k = k_1 \mathbf G_1 + k_2 \mathbf G_2$ for reciprocal lattice vectors $\mathbf G_{1,2}$ of graphene). Panels {\bf c, d}: Cuts along the dashed white lines of the phase diagram in Fig.~\ref{fig:Schematics}{\bf b,c}, displaying $\rho(0)$ and $\mathcal I_M(q=2,L)$ [Eq.~\eqref{eq:IPR}]. These illustrate sequences of semimetallic and metallic transitions concomitant with momentum space delocalization (see Fig.~\ref{fig:WF}). Panels {\bf e} - {\bf j}: The twist dispersions illustrate the difference between semimetallic phases ({\bf e,f,i,j}) and the metallic phase ({\bf g,h}) as well as the remarkably reduced bandwidths (note the reduced scale).
The 2D SOC (cTBG) data were obtained for $Q = 2\pi F_{n - 2}/F_n$ ($\theta = 2\arcsin(\sqrt{3} F_{n-5}/[2F_n])$) at $L = 144$ ($L=377$) and KPM expansion order $N_C = 2^{12}$ ($N_C = 2^{13}$) in the calculation of the DOS while $L = 233$ in panels {\bf e} - {\bf j}.}
\label{fig:DOSandDISP}
\end{figure*}

\subsection{Effective models.}
A variety of effective models (defined in {Supplementary Note 1}) illustrate our proposal.
Here, we focus on two models: a 2D tight-binding Hamiltonian of ``perfect'' spin-orbit coupling (SOC) on a square lattice and a lattice model of TBG at moderate twist angles ($\theta \approx 9^{\circ}$) in the chiral limit (cTBG)  that disallows interlayer tunneling between equivalent sub-lattices~\cite{Tarnopolsky2019} (we fix the bare lattice spacing to unity and $\hbar=1$).
Nonetheless, our main conclusions also apply to TBG beyond the chiral limit for similar twist angles. {(Here, we consider the chiral limit of TBG as it provides the clearest presentation of magic-angle criticality but such a transition can also be shown to persist in the full TBG model. This study will appear elsewhere.)}
The SOC model is given by a hopping $\hat T_{\rm SOC} = t/2\sum_{\vec r,\mu} (i c_{\vec r}^\dagger \sigma_\mu c_{\vec r + \hat \mu} + \mathrm{h.c.})$ and a quasiperiodic potential
\begin{equation}
\hat V_{\rm SOC} = W \sum_{\vec r,\mu = x,y}    \cos(Q r_\mu + \phi_{\mu}) c_{\vec r}^\dagger c_{\vec r},
\label{eq:H0}
\end{equation}
where the $\sigma_\mu$ are Pauli matrices, $c_\br$ are two-component annihilation operators, $t$ is the hopping strength, and $\phi_\mu$ is the offset of the origin.
The lattice model that captures the low-energy theory of cTBG at incommensurate twist angles contains $\hat T_{\rm cTBG}$ that describes nearest neighbor hopping (amplitude $t=\unit[2.8]{eV}$) on the honeycomb lattice.
The interlayer tunneling in the chiral limit is given by
\begin{multline}
\hat V_{\rm cTBG} = W \sum_{\br,\mu}\bigg[ \cos(\mathbf q_\mu \cdot\tfrac{\br + \br_\mu}2 + \phi_\mu)c_{1A\br_\mu}^\dagger c_{2B\br} -
\\
\sum_{n=1}^6 \tfrac{(-1)^n}{3\sqrt{3}}  \sin(\mathbf q_\mu \cdot\tfrac{\br + \br_{\mu n}}2 + \phi_\mu) c_{1A\br_{\mu n}}^\dagger c_{2B\br} + (A\leftrightarrow B)\bigg] + \mathrm{h.c.}
\label{eq:tbg}
\end{multline}
where $c_{lA/B\mathbf r}$ annihilates an electron on layer $l$, sublattice $A/B$, and position $\br$. The index $\mu=1,2,3$ labels nearest neighbors such that $\br_1 -\br = (0,1)$ [$\mathbf q_1 = k_\theta(0,-1)$] with $\br_{\mu}-\br$ [$\mathbf q_\mu$] being $120^\circ$ rotations of the previous vector. The positions $\br_{\mu n} = \br_\mu + \mathbf a_n$ where $\mathbf a_1 = (\sqrt{3}/2,3/2)$ and each subsequent $\mathbf a_n$ is a $60^\circ$ rotation of the last (i.e.\ labeling nearest neighbors on the triangular Bravais lattice). Last, $|\mathbf q_\mu| = k_\theta = \tfrac{8\pi}{3\sqrt{3}} \sin(\theta/2)$ encodes the twist angle, and $\sum_\mu \phi_\mu = 0$ to satisfy $C_6$ symmetry.
Typically the offsets $\phi_\mu$ in either model are averaged over.
The kinetic part $\hat T_{\rm SOC}$ ($\hat T_{\rm {cTBG}}$) has a momentum-space dispersion with four (two) Dirac nodes and a velocity $v_0=t$ ($v_0=3t/2$), see Fig.~\ref{fig:DOSandDISP}{\bf a} (\ref{fig:DOSandDISP}{\bf b}) inset.
Returning to Eq.~\eqref{eq:H0}, we see that $Q$ replaces the role of the twist angle in Eq.~\eqref{eq:tbg}; unless otherwise stated, we highlight incommensurate effects by taking $Q = 2\pi/\varphi^2$ ($\theta = 2\arcsin(\sqrt{3}/2\varphi^5)\approx 8.96^\circ)$
where $\varphi$ is the golden ratio, and in numerical simulations we employ rational approximants $Q_n\equiv 2\pi F_{n-2}/F_n$ ($k_\theta$ is approximated using continued fractions, see {Supplementary Note 1} for details) where the system size $L=F_n$ is given by the $n$th Fibonacci number~\cite{PixleyGopalakrishnan2018}. Other values, in particular smaller $\pi-Q$ and $\theta$, are discussed in the Supplementary Information and below.
In the low-energy approximation this model is identical to the continuum model studied in Ref.~\cite{Tarnopolsky2019} where exact flatbands are uncovered and explained; this makes this model ideal to study incommensurate effects on the lattice.

In addition to Eqs.~\eqref{eq:H0} and \eqref{eq:tbg} we have studied a multitude of other $d$-dimensional models in an incommensurate potential: the $\pi$-flux model and the honeycomb model in 2D, a 3D variant of Eq.~\eqref{eq:H0} (studied previously in Ref.~\cite{PixleyGopalakrishnan2018}), and a 1D long range hopping model with a power-law dispersion $E = -t\; \text{sign}(\cos k) \vert \cos k\vert^\sigma$ with $\sigma <1$
\cite{GaerttnerRadzihovsky2015}---in this 1D case, $v$ is not a velocity (details on 1D case can be found in the last part of {Supplementary Note 2}).
Each of these models generates flat bands and magic-angle physics similar to TBG.
Importantly, these semimetallic 2D Dirac points have been realized in cold atomic setups using either a honeycomb optical lattice~\cite{TarruellEsslinger2012,JotzuEsslinger2014} or artificial gauge fields~\cite{AidelsburgerGoldmann2014,HuangJing2016,WuZhang2016}, whereas the 1D model we consider can be realized using trapped ions~\cite{RichermeMonroe2014}.
The 3D variant of Eq.~\eqref{eq:H0} is theoretically possible to implement~\cite{SunDasSarma2011,Jiang2012,DubcekBuljan2015}, but has not been experimentally realized yet.
In each of these experimental setups, quasiperiodic potentials can then be realized, e.g. by additional lasers \cite{SchreiberBloch2015}, programmable potentials \cite{SmithMonroe2016}, or a digital mirror device \cite{WeitenberKuhr2011}. Alternative emulators of Dirac semimetals can also be realized in metamaterials, e.g.~in topolectrical circuits~\cite{LeeThomale2017}  or in arrays of electromagnetic microwave resonators~\cite{Peterson2018}. Quasiperiodicity can then be encoded through the spatial modulation of the electrical circuit elements.

\subsection{Single-particle spectrum and velocity renormalization.}\label{sec:incommensuratespectrum}
We first discuss the spectral characteristics of magic-angle semimetals probed through the DOS, 
defined as $\rho(E) = 1/N_H \sum_i \delta(E-E_i)$ where $E_i$ is the $i$th eigenenergy
and $N_H$ is the size of the single particle Hilbert space.
At weak quasiperiodic modulation the semimetal is stable, i.e.~$\rho(E)$ vanishes at zero energy with the same power law as in the limit of $W=0$, while hard spectral gaps and van Hove
singularities
develop at finite energy. For Weyl and Dirac Hamiltonians the low-$|E|$ DOS obeys $\rho(E)\sim v^{-d}|E|^{d-1}$, and as $W$ increases, the $(d-1)$st derivative of the DOS [$\rho^{(d-1)} (0)\propto 1/v^d$] increases, see Fig.~\ref{fig:DOSandDISP}{\bf a}, {\bf b} for the model in Eqs.~\eqref{eq:H0} and \eqref{eq:tbg}, respectively.
These weak coupling features may be understood at the level of perturbation theory.

We find that gaps appear at finite energy due to the hybridization around Dirac nodes a distance $Q$ (or $\sqrt{3} k_\theta$) away in momentum space,
see the 
insets in Fig.~\ref{fig:DOSandDISP}{\bf a} and {\bf b}, inset.
For the SOC (cTBG) model, this process ``carves out'' a square {(hexagon)} around each Dirac cone which contains $2 [(\pi-Q)L/2\pi]^2$ ($4[3 \sqrt{3} k_\theta L / 4\pi]^2$) states.
For a given \emph{incommensurate} $Q$ or $\theta$,
there is an infinite sequence of relevant orders in perturbation theory that produce quasi-resonances and open up gaps near zero energy, forming minibands; this is in contrast to the \emph{commensurate} case when this sequence is finite.
For example, for 2D SOC and $Q=2\pi/\varphi^2$, the infinite sequence is given by half the even Fibonacci numbers $F_{3n}/2$, which is the sequence $1,4,7,72,305,\ldots$ (see {Supplementary Note 3}). 
In the incommensurate limit, near the magic-angle transition this sequence of gaps produces a corresponding sequence of minibands, shown in Fig.~\ref{fig:Schematics}{\bf d} for the second, third, and fourth.
We explore the effect of this sequence of minibands using
superlattices in Sec.~\ref{sec:Commensurate}.

Similar to TBG, the renormalization of the velocity in the 2D SOC model can be analytically determined using fourth-order perturbation theory (details in {Supplementary Note 3})~\cite{BistritzerMacDonald2011}.
In terms of the dimensionless coupling constant $ \alpha = {W}/[{2t \sin(Q)}]$ for Eq.~\eqref{eq:H0} this yields
\begin{equation}
\frac{v(W)}{v(0)} =  \frac{1 - 2 \alpha^2 [1-\cos(Q)] + \alpha^4 \frac{4 - 5 \cos(Q)+6 \cos(2Q)}{\cos(Q)}}{1 + 4 \alpha^2 + \alpha^4 \lbrace 16 + [2+1/\cos(Q)]^2\rbrace}. \label{eq:vpert}
\end{equation}
The root of the numerator captures the first magic-angle transition line well when $Q>\pi/2$,
see Fig.~1{\bf b}, independently of whether $Q$ is commensurate or incommensurate.
To describe additional magic-angles, as observed in our numerical data in Fig.~\ref{fig:Schematics}{\bf b},{\bf c}, higher order perturbation theory is required.
For reentrent semimetallic phases, Eq.~\eqref{eq:vpert} indicates the reversal of the Berry phase, consistent with the inversion of miniband states in 3D~\cite{PixleyGopalakrishnan2018}.
In each model we have considered for $d>1$, we have found that the perturbative expression for the velocity (summarized in {Supplementary Table 1}) has  a magic-angle condition where the velocity vanishes.

As the magic-angle is approached, higher perturbative corrections become relevant.
To go beyond perturbation theory, we compute the DOS using the numerically exact kernel polynomial method (KPM), on sufficiently large system sizes across a range of models of various dimensions.
At a critical $\alpha  = \alpha_c \sim 1$ the DOS becomes {non-analytic} and a metallic spectrum with finite $\rho(0)$ develops for $\alpha > \alpha_c$, see Fig.~\ref{fig:DOSandDISP}{\bf c,d} (for cTBG $\alpha = \tfrac{W}{2t\sin(3k_\theta/4)}$).
In particular, for $d>1$ and fixed $Q$ or $\theta$, $\rho(E) \sim |W - W_c|^{-\beta} |E|^{d - 1}$ implying the velocity $v(W) \sim \vert W-W_c\vert^{\beta/d}$.
Surprisingly, we find $\beta\approx 2$ in each model  investigated above 1D (see {Supplementary Note 2})~\cite{PixleyGopalakrishnan2018},
indicating that this exponent is universal.
In 1D this magic-angle effect also exists but is modified by the form of the dispersion such that $\rho(E) \sim |W - W_c|^{-\beta} |E|^{1/\sigma -1}$, and for the case $\sigma=1/3$ we find $\beta = 4.0\pm 0.8$.

This velocity renormalization is the manifestation of the aforementioned reconfiguration of the band structure and the appearence of a sequence of minibands. Of course, broken translational symmetry precludes a standard bandstructure of dispersive Bloch waves.  In Fig.~\ref{fig:DOSandDISP} {\bf e}-{\bf j} we therefore illustrate this reconfigured bandstructure, at a fixed rational approximant, in the form of
the twist dispersion (obtained by exact diagonalization in the presence of twisted boundary conditions) along high symmetry lines for the models defined in Eqs.~\eqref{eq:H0}, \eqref{eq:tbg}. We performed the analogous analysis for a multitude of models and plotted the velocity $v(W)$ near the semimetallic touching points in Fig.~\ref{fig:Schematics}{\bf a}. The velocity $v(W)$ as determined by computing the twist dispersion agrees with the calculation of $\rho^{(d-1)}(0)$, see {Supplementary Note 2}.

\subsection{Critical single-particle wave functions}

Magic-angle semimetals are intimately linked to the physics of Anderson transitions in momentum space; this is captured by the eigenfunctions near 
the Dirac node energy, $E=0$~\cite{PixleyGopalakrishnan2018}. 

\begin{figure*}[t]
\includegraphics[width=.95\textwidth]{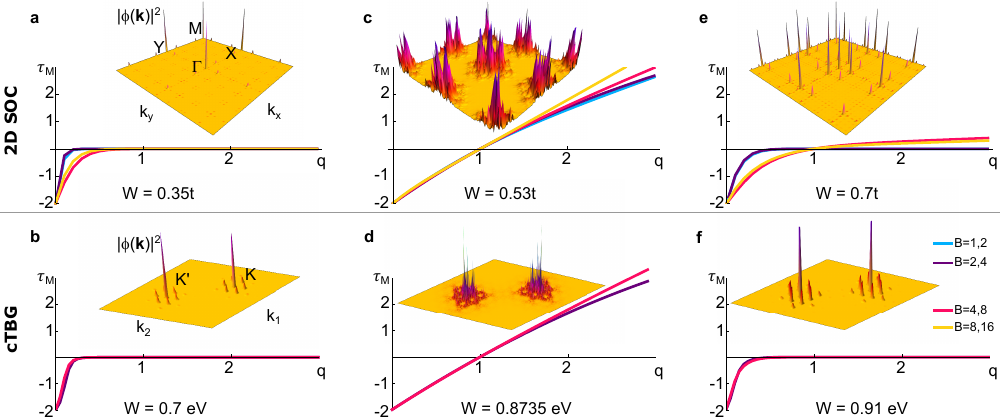}
\caption{{\bf Eigenstate transition as manifested in momentum space wave functions at the Dirac node energy $E=0$.} Panels {\bf a} - {\bf f}: Wave function characteristics as described by the scaling exponent $\tau_M(q)$ averaged over 100 random phases and twisted boundary conditions. For $W < W_{c}$ and $W> W_{c}'$ the wave functions are ballistic [with a frozen $\tau_M(q)$] while for $W_{c} < W < W_{c}'$ they are critical in momentum space [$\tau_M(q)$ is weakly non-linear in $q$].  Inset of {\bf a} - {\bf f}: corresponding momentum space wavefunctions.
The 2D SOC (cTBG) data were obtained for $Q = 2\pi F_{n - 2}/F_n$ ($\theta = 2\arcsin(\sqrt{3} F_{n-5}/[2F_n])$)  at $L = 144$ ($L=377$).}
\label{fig:WF}
\end{figure*}

We compute the low-energy wavefunctions using
Lanczos for large $L$ reaching up to $L= 377$ and $610$ in the cTBG and SOC models, respectively.
Qualitatively, we find that the structure of the wave functions in the semimetallic phase is stable and adiabatically connected to the ballistic $W = 0$ limit, with isolated ballistic spikes in momentum space, see Fig.~\ref{fig:WF}{\bf a},{\bf b}.
In contrast, the form of the wave functions is completely different in the metallic state, see Fig.~\ref{fig:WF}{\bf c},{\bf d}, as it appears delocalized both in momentum and real space {with non-trivial structure} (see details in {Supplementary Note 5}). 
Finally, in the reentrant semimetal, the wave functions are again ballistic, see Fig.~\ref{fig:WF}{\bf e},{\bf f}.
Crucially, in all models that we studied, the positions of the transitions in the spectral properties of the DOS coincide with the transitions of the wave functions characteristics within numerical resolution, see Figs.~\ref{fig:DOSandDISP}{\bf c},{\bf d}.

In order to quantify the eigenstate QPTs of the wave functions, we generalize the multifractal wave function analysis~\cite{EversMirlin2008} to momentum space.
We define the inverse participation ratio of the energy eigenstates in momentum  space~\cite{PixleyGopalakrishnan2018} $\psi_E(\vec k)$ at a given energy $E$
\begin{equation}
\mathcal I_M (q,L) = \sum_{\vec k} \vert \psi_E(\vec k) \vert^{2q} \,
\sim L^{-\tau_M(q)}. \label{eq:IPR}
\end{equation}
We can now apply properties of the scaling exponent $\tau_M(q)$, typically used to analyze real space localization, to momentum space.
It monotonically increases [obeying $\tau_M(0) = - d$ and $\tau_M (1) = 0$] and distinguishes delocalized wave functions 
[$\tau_M(q) = d(q-1)$]
from exponentially localized peaks [$\tau_M(q>0) = 0$] and critical states with non-linear ``multifractal'' $\tau_M(q)$.
A variant of multifractal states, which are called ``frozen,'' display $\tau_M(q>q_c) = 0$ for a given $q_c \in (0,1]$; their peak height is system size independent, as in standard localized states, but show multifractal correlations in their tails~\cite{EversMirlin2008}.
We employ the standard binning technique (varying the binning size $B$) to numerically extract the scaling exponents $\tau_M(q)$ in systems of a given finite size, see {Supplementary Note 5} for details.

Focusing on $q=2$, as shown in Fig.~\ref{fig:DOSandDISP}{\bf c}, {\bf d} for the SOC and cTBG models, respectively, the momentum space wavefunction at the Dirac node energy delocalizes upon crossing the magic-angle in the incommensurate limit.
The momentum space delocalization can 
not occur in
the commensurate case; Bloch's theorem ensures the existence of states with well defined (i.e.~well localized) crystalline momenta. For example,
consider Eq.~\eqref{eq:H0} in the commensurate limit where $Q/2\pi = a/b$ ($a$ and $b$ are coprime integers). 
In this case, $\mathcal I_M (q,L)$ is bounded from below by $1/b^{d(q-1)}$ and hence $\tau_M(q)=0$ in the thermodynamic limit $L/b \rightarrow \infty$ preventing momentum space delocalization (see {Supplementary Note 5}). In contrast, we here numerically access the incommensurate limit using finite size scaling of rational approximants corresponding to $L = b \rightarrow \infty$.

The scaling analysis of $\mathcal I_M(q,L)$ at the energy of the Dirac node $E=0$, presented in Figs.~\ref{fig:WF}{\bf a}-{\bf f} for Eqs.~\eqref{eq:H0} and~\eqref{eq:tbg},
demonstrates three phases of distinct wavefunction structures in momentum space.
A frozen spectrum $\tau_M(q)$ occurs in the two semimetal regimes. 
In sharp contrast, the function $\tau_M(q)$ unfreezes in the metallic phase with finite $\rho(0)$.
Surprisingly, throughout the metallic phase the spectra appear to be weakly multifractal in both momentum and real space ({Supplementary Note 5}), we find
for the SOC model that
$\tau_M(q) \approx 2(q-1)-0.25q(q-1)$ and for the cTBG model we obtain $\tau_M(q) \approx 2(q-1)-0.15q(q-1)$ (in the region $|q|<1$ and within the limits of our numerical precision)
in Fig.~\ref{fig:WF}{\bf c},{\bf d}, which are both \emph{non-linear} in $q$.
The observation of similar behavior in all models that we investigated (as listed in {Supplementary Note 1}) corroborates the interpretation of the magic-angle phenomenon 
in the incommensurate limit
as one of eigenstate quantum criticality and generalizes the quasiperiodic 3D Weyl semimetal-to-diffusive metal QPT~\cite{PixleyGopalakrishnan2018} to arbitrary dimensions.
In two dimensions we do not find any signatures of diffusion (consistent with the marginality of two dimensions \cite{AbrahamsRamakrishnan1979,DevakulHuse2017}) and in one dimension the semimetal transitions directly to an Anderson insulator (shown in {Supplementary Note 2}).
Lastly, when $d>1$ and $W$ is substantially larger than the magic-angle  transition, all investigated models undergo Anderson localization in real space (e.g. at $W > 1.75t$ in the case of the 2D SOC model at $Q = 2\pi/\varphi^2$).

\subsection{Commensurate superlattices and Hubbard models.}
\label{sec:Commensurate}

So far, our analysis regarded non-interacting magic-angle semimetals in the strict incommensurate limit.
We now turn to the interparticle interaction term $\hat U$ in the Hamiltonian in Eq.~\eqref{eqn:genH} in commensurate superlattices.
In order to illustrate how the appearance of flatbands enhances correlations, we construct a series of emergent Hubbard models near the magic-angle transition for Eq.~\eqref{eq:H0} at $\phi_\mu = \pi/2$ supplemented by
\begin{equation}
  \hat U_{SOC} = U \sum_{\mathbf{r}} n_{\mathbf{r},\uparrow}n_{\mathbf{r},\downarrow},
  \label{eq:Hubbard}
\end{equation}
with $n_{{\bf r} \sigma}=c_{{\bf r} \sigma}^{\dag}c_{{\bf r} \sigma}$.
In contrast to the previous discussion, we take commensurate approximations in order to build translationally invariant Hubbard models.
In particular, we still use the rational approximants $Q_n = 2\pi F_{n-2}/F_n$, only now we take the size of the system $L = m F_n$ for some integer $m$, effectively taking the thermodynamic limit in $L$ before the limit of quasiperiodicity $Q_n \rightarrow Q$.
This is reminiscent of moir\'e lattices used to model TBG, and similarly, we can unambiguously define a supercell of size $\ell = F_n$  and isolate bands in $k$-space.

\begin{figure}
   \includegraphics[width=0.45\textwidth]{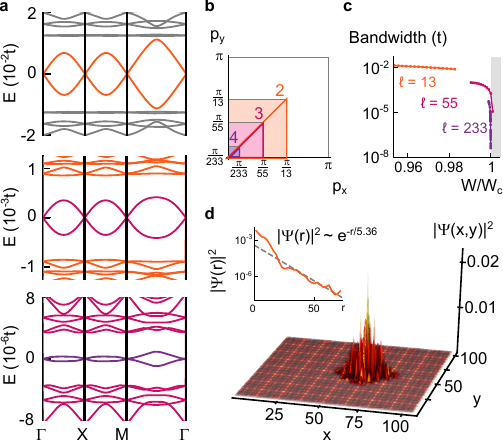}
  \caption{{\bf Supercell analysis and Wannier functions.} The color coding matched across {\bf a}--{\bf c} (and Fig.~\ref{fig:Schematics}{\bf d}) indicates the 2nd (orange), 3rd (maroon), and 4th (purple) minibands. {\bf a}. The dispersion of Eq.~\eqref{eq:H0} in the mini-Brillouin zone for superlattices $(\ell,W) = (13, 0.5), (\ell, W) = (55, 0.5244), (\ell,W) = (233, 0.5244)$ (from top to bottom); this illustrates successive emergence of minibands (from top-to-bottom) as a consequence of consecutive downfoldings of the Brillouin zone. 
  {\bf b}. The corresponding mini-Brillouin zones (logarithmic scale). {\bf c}. The dramatic reduction in bandwidth near the critical point for each miniband. {\bf d}. For $(\ell, W)=(13, 0.5)$ and $L=104$, computed Wannier function $\psi(x,y)$ that is sitting upon the local density of states $\rho_{\mathrm{band}}(\mathbf r) = \sum_{n} |\braket{\mathbf r|E_n}|^2$ (shown as a density plot) for eigenstates of the (orange) band $\ket{E_n}$, on a $104\times 104$ lattice.
  (Inset). The exponential localization of the Wannier state. }
  \label{fig:Spectrum}
\end{figure}

In particular, these bands are intimately related to the hierarchy of minibands derived with perturbation theory: 
when $\ell = F_{3a + b}$ for integers $a$ and $b=1,2$, the gap for the 
central band opens at order $F_{3a}/2$ in perturbation theory (for $\ell = F_{3a}$, the Dirac nodes gap at order $F_{3a}/2$. See {Supplementary Note 3} for details).
The series of superlattices indicated by $\ell$ correspond to the sequence of gap openings in Sec~\ref{sec:incommensuratespectrum} --- making the notion precise --- with (downfolded) Brillioun zones depicted in Fig.~\ref{fig:Spectrum}{\bf b}. 
Near $W_c$, hard gaps open and the minibands form as 
illustrated in Fig.~\ref{fig:Spectrum}{\bf a} 
for $\ell = 13,55,233$ (respectively, the 2nd, 3rd, and 4th minibands). 
We conjecture that all of these minibands (as $\ell\rightarrow\infty$) achieve gaps near $W_c$ as evidenced by Fig.~4{\bf a,c} in the incommensurate limit, indicating something akin to the singular continuous spectrum of the Aubry-Andr\'e model at criticality~\cite{HiramotoKohmoto1992}.
Furthermore, the central band becomes flatter, as indicated by the reduction in bandwidth seen in Fig.~\ref{fig:Spectrum}{\bf c} which we track until the dispersion loses its semimetallic character.

We exploit this miniband formation and the existence of hard gaps to build symmetric Wannier functions in the semimetallic regime, see Fig.~\ref{fig:Spectrum}{\bf d}.
To build the Hubbard models, we perform approximate joint diagonalization on the position operators ($\hat x_\mu$)  projected (with projection operator $P$) onto a given band  $\hat X^{\mathrm{MB}}_{\mu} \equiv P \hat x_{\mu} P$ in order to determine the Wannier states~\cite{MarzariVanderbilt2012} (for details and code, see {Supplementary Note 6}).
We have checked that not only are the computed Wannier states exponentially localized to numerical precision (Fig.~\ref{fig:Spectrum}{\bf d}, inset), but that they are also symmetric.
Therefore, the minibands formed from the SOC model and pictured in Fig.~\ref{fig:Spectrum} are \emph{not} topological~\cite{brouderExponentialLocalizationWannier2007}, fragile~\cite{canoTopologyDisconnectedElementary2018,poFragileTopologyWannier2018} or otherwise.

As a clear example, when $W=0.5t$ and $(\ell,m) = (13,8)$, we see a clear band around $E=0$ in Fig.~\ref{fig:Spectrum}{\bf a}, and we find Wannier centers in a well defined grid (Fig.~\ref{fig:Spectrum}{\bf d}, main panel) corresponding to exponentially localized Wannier states (Figs.~\ref{fig:Spectrum}{\bf d}, inset).
The projected Hamiltonian has the approximate form of Eqs.~\eqref{eq:H0} and~\eqref{eq:Hubbard} with a renormalized $U_\mathrm{eff}$, $t_\mathrm{eff}$ and $W_{\rm eff} = 0$.
With this approach, we can identify successive gaps leading up to the metallic transition from either side along with dramatic enhancements of interactions,  which reach up to a massive $U_\mathrm{eff}/t_\mathrm{eff}  \sim 4100 U/t$ for the fourth miniband with supercell $\ell = 377$, as shown in Fig.~\ref{fig:Schematics}{\bf d}.
This can also been shown analytically using a one step renormalization group calculation, which yields the divergence $U_\mathrm{eff}/t_\mathrm{eff} \sim U (1/\ell)^{d-1} Z^2/v\sim 1/|W-W_c|$, ($\sqrt Z$ is the wave function renormalization), as shown in detail in {Supplementary Note 3}.
Due to finite size, the apparent location of $W_c$ can artificially shift, therefore in Fig.~\ref{fig:Schematics}{\bf d} we use $W_c = \tilde W_c \frac{\sin Q}{\sin Q_n}$ where $\tilde W_c$ is the transition point when $n\rightarrow\infty$.
In {Supplementary Figure 16} we present the data for a large set of $(\ell, m)$ corroborating our findings.

Away from $E=0$, nearly flat (semimetallic) bands can form well before the magic-angle transition with similarly large $U_\mathrm{eff}/t_\mathrm{eff}$, see Fig.~\ref{fig:Spectrum}{\bf a}.
In very close proximity to the transition, multi-orbital Hubbard models appear (see {Supplementary Note 6}).

\subsection{Experimental cold atomic realization}

\begin{figure}
\includegraphics[width=0.45\textwidth]{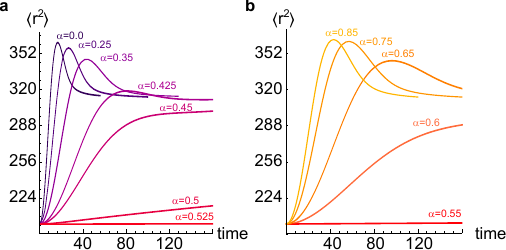}
\caption{{\bf Boltzmann wave packet spreading.} Spreading of the mean square radius $\langle r^2 \rangle = \sum_{\bf r} {\bf r}^2 \rho({\bf r})$ of the particle density $\rho({\bf r})$ as a function of time in units of the inverse hopping rate $1/t$ (panel {\bf a}: $\alpha < \alpha_c$, panel {\bf b}: $\alpha > \alpha_c$). Here, we consider the interacting 2D SOC model, Eqs.~\eqref{eq:H0} and~\eqref{eq:Hubbard}, and we employ Eq.~\eqref{eq:vpert} to incorporate the magic angle effect (occuring at $\alpha_c \approx 0.53$ in this approximation) into a semianalytical hydrodynamic treatment. The initial steady state at finite temperature is defined by a particle [energy] density $\rho (\textbf{{r}}) =  {e^{- \textbf{{r}}^2/[2\xi^2]}}/{\xi^2}$ [$\rho_E(\textbf{{r}}) = {v_0}\left (1 + 3 e^{- \textbf{\textit{r}}^2/\xi^2} \right)/{\xi^3}$], with $v_0 \equiv v(\alpha=0)$ is the bare velocity and we chose $\xi = 4$ for the initial spread of the density profile. The hydrodynamic equations were numerically solved in the presence of an onsite repulsion $U(\alpha = 0) = 0.025 t$ and Umklapp scattering rate $1/\tau(\alpha = 0) = 0.0075 t$ (see {Supplementary Note 4}).  }
\label{fig:Boltzmann}
\end{figure}

All sufficient ingredients
for emulating
magic-angle phenomenon
are available in ultra-cold atomic gas and metamaterial~\cite{Rechtsman2008,Peterson2018} experiments.
In particular for ultra-cold atomic gases, to probe fermionic strong correlations, the atomic species ${}^{40}$K and ${}^6$Li are prime candidates; we estimate that the underlying lattice can be relatively shallow (around 8 lattice recoil energies), and need temperatures relative to the Fermi temperature (of the entire gas) $T/T_{\mathrm F}\approx 0.25$ to ensure fermion population fills but does not exceed the first miniband.
To see large correlations, trap sizes should accommodate at least roughly $30\times 30$ optical lattice sites.
In addition to any spectroscopic measurements that probe the density of states (e.g. radiofrequency spectroscopy~\cite{Gaebler-2010}), we propose and demonstrate (in more detail in {Supplementary Note 4}) that the analysis of wavepacket dynamics is an indicator of magic-angle physics. In the absence of interactions, we numerically predict a non-monotonic spreading of the wave function for increasing $W$ (see {Supplementary Figure 10}) in the regime with multiple magic angles. We have also studied the
interacting model in the hydrodynamic regime
by using a generalization of the Boltzmann kinetic equation~\cite{SchneiderRosch2012} (see details in {Supplementary Note 4}). Its solution confirms the drastic decrease of the expansion velocity and a substantial enhancement of diffusive dynamics near the magic angle, see Fig.~\ref{fig:Boltzmann}. The observation of these effects is possible within experimentally realistic observation time of $50 t^{-1}$ ($\sim \!\! 10$--$100$ ms).
Moreover, our work {demonstrates} an experimental protocol for realizing strong correlations by first cooling the gas to quantum degeneracy and then applying a quasiperiodic potential to create flat bands without the need to cool the system in a Mott insulator phase or load the atoms into a flat band.

\section{Discussion} 
In summary, we introduced a class of magic-angle semimetals
and
demonstrated the general appearance of a single-particle quantum phase transition in the incommensurate limit at which, simultaneously, (i) the kinetic energy vanishes universally, (ii) a non-zero density of states appears at zero energy, and (iii) the wave functions display delocalization and multifractality in momentum space.
In  
the presence of interactions we demonstrated that this eigenstate criticality leads to a strongly correlated Hubbard model by computing Wannier states on a superlattice. Lastly, we presented a detailed discussion of an experimental realization in cold atomic quantum emulators.

Regarding experimentally realized twisted graphene heterostructures at much smaller twist angles than we have considered here ($\theta \approx 1.1^{\circ}$), it has not been obvious whether incommensuration is an important ingredient~\cite{Mele-2010}.
Quasiperiodic effects rely upon weakly detuned processes at which the total transferred momentum wraps the  Brillouin zone.
In contrast, the momentum transfer induced by scattering off a small angle superstructure is minute.
Therefore---it is often concluded---both effects of incommensurability and intervalley scattering are negligible as processes in higher order perturbation theory. As our numerics demonstrate, this results in the suppression of the width of the metallic sliver in Fig.~\ref{fig:Schematics}{\bf b}, {\bf c} that scales like $W_c' - W_c \sim \theta^3$, making observing such a metallic phase exceedingly difficult at small twist angles.
Nonetheless, we expect Anderson delocalization in momentum space even at small twist angles.
This is because this physics is dominated by rare resonances (as manifested in the locator expansion~\cite{Scardicchio2017}) and controlled by $\alpha$, while perturbative processes are parameterized by $W/t$ and are therefore small.
Furthermore, the contiguous phase boundary in Fig.~\ref{fig:Schematics}{\bf b}, {\bf c}
may imply
that the physics of small angles directly connects to large, incommensurate twists~\cite{PalSpitz2018,Yao2018,AhnAhn2018}.
However, within present day numerics, we cannot exclude that this boundary of eigenstate phase transitions terminates at a finite, small angle, which would imply the existence of a critical Anderson delocalization end point in Fig.~\ref{fig:Schematics}{\bf b,c}.
The coexistence of finite DOS with other features of this phase at larger angles suggests that the phase extends to $\theta \rightarrow 0$ ($Q\rightarrow\pi$), but an end-point is appealing as it would establish a theoretical paradigm of quasiperiodic Anderson tricriticality.
Any rational approximant or commensurate angle truncates the infinite sequence of resonances and minibands which leads to a rounding of the QPT (akin to finite size effects in usual transitions) and the absence of momentum space delocalization.
The amplified interactions due to flat bands and an enhanced DOS occur for both incommensurate and commensurate cases as Fig.~\ref{fig:Schematics}{\bf d} demonstrates.
This enhancement coupled with eigenstate quantum criticality in the incommensurate limit  characterizes magic-angle semimetals, including twisted bilayer graphene at moderate twist angles.

\textit{Note added}: While this manuscript was under consideration following its announcement in arXiv:1809.04604, independent proposals to simulate twistronics in cold atoms appeared and were published in ref.~\cite{gonzalez2019cold,Salamon2020}

\section{Methods}

\subsection{{ Numerical Methods.}}

The numerical methods used are the kernel polynomial method \cite{weisseKernelPolynomialMethod2006} for the density of states, exact diagonalization and Lanczos for eigenstates, approximate joint diagonalization for Wannier functions \cite{marzariMaximallyLocalizedWannier2012}, and numerical partial differential equation solvers for the Boltzmann kinetics.
These methods are explained in context in the Results section with additional details in the supplementary information, particularly {Supplementary Notes 2, 4, 5 and 6}.

\section*{Data Availability}
The data sets generated during and/or analysed during the current study are available from the corresponding author on reasonable request.

\section*{Code Availability}
The numerical code used to generate this data is available upon request. For Wannier functions, the code used for the approximate joint diagonalization can be found here \href{https://github.com/jhwilson/AJD.jl}{https://github.com/jhwilson/AJD.jl}.

\section*{Acknowledgements}
We thank I. Bloch, P.-Y. Chang, P. Coleman, B. J. DeSalvo, M. Foster, Y. Komijani, G. Pagano, A. M. Rey, M. Sch\"utt, I. Spielman, and D. Vanderbilt for useful discussions. We also thank S. Gopalakrishnan and D. Huse for collaborations on related work and for insightful discussions.
J.H.W. and J.H.P. acknowledge the Aspen Center for Physics where some of this work was performed, which is supported by National Science Foundation Grant No. PHY-1607611.
J.H.P. is partially supported by the Air Force Office of Scientific Research under Grant No.~FA9550-20-1-0136.
E.J.K. was supported by the U.S. Department of Energy, Basic Energy Sciences, grant number DE-FG02-99ER45790.
This work was supported by the Caltech Institute for Quantum Information and Matter, an NSF Physics Frontiers Center with support of the Gordon and Betty Moore Foundation and the Air Force Office for Scientific Research (J.H.W.).
Y.-Z.C. was supported in part by a Simons Investigator award to Leo Radzihovsky and in part by the Army Research Office under Grant Number W911NF-17-1-0482.
The views and conclusions contained in this document are those of the authors and should not be interpreted as representing the official policies, either expressed or implied, of the Army Research Office or the U.S. Government.
The U.S. Government is authorized to reproduce and distribute reprints for Government purposes notwithstanding any copyright notation herein.
The authors acknowledge the Beowulf cluster at the Department of Physics and Astronomy of Rutgers University, The State University of New Jersey, and the Office of Advanced Research Computing (OARC) at Rutgers, The State University of New Jersey (http://oarc.rutgers.edu) for providing access to the Amarel cluster and associated research computing resources that have contributed to the results reported here.

\section*{Statement of Competing Interest}
The Authors declare no competing financial or non-financial interests.

\section*{Author Contribution}
E.~J.~K., J.~H.~W., and J.~H.~P.~designed the research, Y.~F., E.~J.~K., J.~H.~W., YZ.~C, and J.~H.~P.~performed the research, and Y.~F., E.~J.~K., J.~H.~W., YZ.~C, and J.~H.~P. wrote the paper. Y.~F., E.~J.~K., and J.~H.~W.\ contributed equally.



\section*{Supplementary material}

# Magic angle semimetals
# SUPPLEMENTAL MATERIAL

Yixing Fu,[1, *] E. J. König,[1, *] J. H. Wilson,[2, 1, *] Yang-Zhi Chou,[3, 4] and J. H. Pixley[1]

[1] *Department of Physics and Astronomy, Center for Materials Theory, Rutgers University, Piscataway, NJ 08854 USA*
[2] *Institute of Quantum Information and Matter and Department of Physics,*
*California Institute of Technology, Pasadena, California 91125 USA*
[3] *Condensed Matter Theory Center and the Joint Quantum Institute,*
*Department of Physics, University of Maryland, College Park, MD 20742 USA*
[4] *Department of Physics and Center for Theory of Quantum Matter,*
*University of Colorado Boulder, Boulder, Colorado 80309, USA*
(Dated: November 25, 2019)

## I. MODELS

In this section we define the models which we analyze.

### A. Semimetals in an incommensurate scalar potential

The tight-binding Hamiltonians of models dubbed "perfect" spin orbit coupling (SOC) are given by

$$\hat{T}_{SOC} = \sum_{\mathbf{r},\mu} [\frac{it}{2} c_{\mathbf{r}}^\dagger \sigma_\mu c_{\mathbf{r}+\hat{\mu}} + \text{h.c.}], \tag{S1}$$

$$\hat{V}_{SOC} = W \sum_{\mathbf{r},\mu} \cos(Qr_\mu + \phi_\mu) c_{\mathbf{r}}^\dagger c_{\mathbf{r}}, \tag{S2}$$

where $t$ is the hopping matrix element, $\sigma_\mu$ are the Pauli matrices, $c_{\mathbf{r}}$ are electron annihilation operators, $Q$ is our quasiperiodic wave vector, $\phi_\mu$ are the offsets of the origin of the potential, and $W$ is the amplitude of the quasiperiodic potential. In the two-dimensional (2D) case $\mu = x, y$ and in the three-dimensional (3D) case $\mu = x, y, z$ and $\mathbf{r}$ takes values in the set of all lattice points on a square (cubic) lattice. The Hamiltonian for the $\pi$-flux model has the same potential term in 2D. The hopping term is modified as follows

$$\hat{T}_\pi = -t \sum_{\mathbf{r},\mu=x,y} [c_{\mathbf{r}}^\dagger e^{iA_\mu(\mathbf{r})} c_{\mathbf{r}+\hat{\mu}} + \text{h.c.}], \tag{S3}$$

where we choose the gauge with $A_x(\mathbf{r}) = \pi/2$ for all sites $\mathbf{r}$ on the square lattice, and $A_y(\mathbf{r}) = -(-1)^{r_x}\pi/2$. For the chosen gauge, periodic boundary conditions require the lattice size in $x$ direction to be even.

The spinless honeycomb (HC) lattice model is given by a Hamiltonian of the form

$$\hat{T}_{HC} = -t \sum_{\mathbf{r}_A, i}[c_A^\dagger(\mathbf{r}_A)c_B(\mathbf{r}_A + \mathbf{d}_i) + \text{h.c.}], \tag{S4}$$

$$\hat{V}_{HC} = W \sum_{\mathbf{r}, \boldsymbol{\delta}_\mu} \cos(Q\mathbf{r} \cdot \boldsymbol{\delta}_\mu + \phi_\mu)c_\mathbf{r}^\dagger c_\mathbf{r}. \tag{S5}$$

The sum over $\mathbf{r}_A$ is over one of the two sub-lattices, while $\mathbf{r}$ is over all points. The index $i$ labels the three nearest neighbors of $\mathbf{r}_A$, and $\mathbf{d}_i$ is the vector from $\mathbf{r}_A$ to its nearest neighbor $i$. The vectors $\boldsymbol{\delta}_\mu$ are a choice of each particular model and for numerics we choose $\boldsymbol{\delta}_1 = \mathbf{d}_1 = (2/3)\hat{x}$ and $\boldsymbol{\delta}_2 = \mathbf{d}_2 = -(1/3)\hat{x} + (1/\sqrt{3})\hat{y}$. The kinetic part of the Hamiltonian for the one dimensional model with power law disperion[S1] is given in momentum space

$$\hat{T}_{1D} = -t \sum_k \text{sgn}[\cos(k)]|\cos(k)|^\sigma c_k^\dagger c_k. \tag{S6}$$

We assume $\sigma < 1$, this expression can be readily Fourier transformed to a tight binding model with long range hopping (LRH). This yields a hopping amplitude

$$t_{ij} \sim -2t[1 - (-1)^{|i-j|}]\sin[\pi(|i-j| - \sigma)/2]\Gamma(1+\sigma)|i-j|^{-(1+\sigma)} \tag{S7}$$

for $|i - j| \gg 1$ and $\Gamma(x)$ is the Gamma function. The potential is

$$\hat{V}_{1D} = W \sum_r \cos(Qr + \phi)c_r^\dagger c_r. \tag{S8}$$

## B. Chiral Twisted Bilayer Graphene

The model we use for twisted bilayer graphene simulates the chiral version of the continuum model[S2–S4].

To write down the full model, we have first that

$$\hat{T}_{cTBG} = -t \sum_{a=1,2} \sum_{\langle ij \rangle} c_{a,\mathbf{r}_i}^\dagger c_{a,\mathbf{r}_j}, \tag{S9}$$

where $\langle ij \rangle$ indicates nearest neighbors on the honeycomb lattice and $a$ labels the two layers. This model has four Dirac nodes: two per layer. Furthermore, for reporting figures, we take $t = 2.8\,\text{eV}$ as it is for graphene.

We then couple the two layers with a quasiperiodically modulated tunnelling to simulate the effect of twisting

$$\hat{V}_{cTBG} = W \sum_\mathbf{r} \sum_j \Big\{ (c_{1,A,\mathbf{r}+\boldsymbol{\eta}_j}^\dagger c_{2,B,\mathbf{r}} + c_{1,B,\mathbf{r}}^\dagger c_{2,A,\mathbf{r}+\boldsymbol{\eta}_j}) \cos[\mathbf{q}_j \cdot (\mathbf{r} + \boldsymbol{\eta}_j/2) + \phi_j]$$

$$- \frac{1}{3\sqrt{3}} \sum_{n=1}^6 (-1)^n (c_{1,A,\mathbf{r}+\boldsymbol{\eta}_j+\mathbf{a}_n}^\dagger c_{2,B,\mathbf{r}} + c_{1,B,\mathbf{r}}^\dagger c_{2,A,\mathbf{r}+\boldsymbol{\eta}_j+\mathbf{a}_n}) \sin[\mathbf{q}_j \cdot (\mathbf{r} + (\boldsymbol{\eta}_j + \mathbf{a}_j)/2) + \phi_j] \Big\} + \text{h.c.}, \tag{S10}$$

where $\mathbf{r}$ is on the triangular Bravais lattice and $\boldsymbol{\eta}_j$ describe the vectors to connect nearest neighbors (B sites to A sites) and they are given by

$$\begin{aligned}
\boldsymbol{\eta}_1 &= (0,1), \\
\boldsymbol{\eta}_2 &= (-\sqrt{3}/2, -1/2), \\
\boldsymbol{\eta}_3 &= (\sqrt{3}/2, 1/2).
\end{aligned} \tag{S11}$$

The vectors $\mathbf{a}_j$ are the six nearest neighbors on the triangular lattice defined by $\mathbf{a}_1 = \boldsymbol{\eta}_1 - \boldsymbol{\eta}_2$, $\mathbf{a}_2 = \boldsymbol{\eta}_1 - \boldsymbol{\eta}_3$, $\mathbf{a}_3 = \mathbf{a}_2 - \mathbf{a}_1$, $\mathbf{a}_4 = -\mathbf{a}_1$, $\mathbf{a}_5 = -\mathbf{a}_2$, and $\mathbf{a}_6 = \mathbf{a}_1 - \mathbf{a}_2$. Last, $\mathbf{q}_j$ are defined by the twist angle

$$\begin{aligned}
\mathbf{q}_1 &= k_\theta(0,-1), \\
\mathbf{q}_2 &= k_\theta(\sqrt{3}/2, 1/2), \\
\mathbf{q}_3 &= k_\theta(-\sqrt{3}/2, 1/2),
\end{aligned} \tag{S12}$$

where $k_\theta = 2k_D \sin(\theta/2)$ for twist angle $\theta$ and $k_D = 4\pi/(3\sqrt{3})$ is the distance from the $\Gamma$ point to $\mathbf{K}$ point.

To show how this reproduces the chiral model of twisted bilayer graphene, we can rewrite the above in $\mathbf{k}$-space

$$
\begin{aligned}
\hat{V}_{\text{cTBG}} = W \sum_{\mathbf{k}} \Big[ & c_{1,\mathbf{k}+\frac{\mathbf{q}_1}{2}}^\dagger \sigma_x c_{2,\mathbf{k}-\frac{\mathbf{q}_1}{2}} e^{i\phi_1} + c_{1,\mathbf{k}+\frac{\mathbf{q}_2}{2}}^\dagger (e^{-i\mathbf{k}\cdot\mathbf{a}_1}\sigma_+ + e^{i\mathbf{k}\cdot\mathbf{a}_1}\sigma_-) c_{2,\mathbf{k}-\frac{\mathbf{q}_2}{2}} e^{i\phi_2} \\
& + c_{1,\mathbf{k}+\frac{\mathbf{q}_3}{2}}^\dagger (e^{-i\mathbf{k}\cdot\mathbf{a}_2}\sigma_+ + e^{i\mathbf{k}\cdot\mathbf{a}_2}\sigma_-) c_{2,\mathbf{k}-\frac{\mathbf{q}_3}{2}} e^{i\phi_3} \Big] f(-\mathbf{k}) + w \sum_{\mathbf{k}} \Big[ c_{1,\mathbf{k}-\frac{\mathbf{q}_1}{2}}^\dagger \sigma_x c_{2,\mathbf{k}+\frac{\mathbf{q}_1}{2}} e^{-i\phi_1} \\
& + c_{1,\mathbf{k}-\frac{\mathbf{q}_2}{2}}^\dagger (e^{-i\mathbf{k}\cdot\mathbf{a}_1}\sigma_+ + e^{i\mathbf{k}\cdot\mathbf{a}_1}\sigma_-) c_{2,\mathbf{k}+\frac{\mathbf{q}_2}{2}} e^{-i\phi_2} + c_{1,\mathbf{k}-\frac{\mathbf{q}_3}{2}}^\dagger (e^{-i\mathbf{k}\cdot\mathbf{a}_2}\sigma_+ + e^{i\mathbf{k}\cdot\mathbf{a}_2}\sigma_-) c_{2,\mathbf{k}+\frac{\mathbf{q}_3}{2}} e^{-i\phi_3} \Big] f(\mathbf{k}), \quad \text{(S13)}
\end{aligned}
$$

where $c_{a,\mathbf{k}} = (c_{aA\mathbf{k}}, c_{aB\mathbf{k}})^T$ with $a$ labeling the layer, A and B labeling the sublattice, and $\mathbf{k}$ is the lattice wave vector. The function $f$ is real-valued and has the form

$$
f(\mathbf{k}) = \frac{1}{2} + \frac{1}{6i\sqrt{3}} \sum_{n=1}^{6} (-1)^n e^{-i\mathbf{k}\cdot\mathbf{a}_n}. \tag{S14}
$$

If we then concentrate near the $\mathbf{K}$ or $\mathbf{K}'$ points ($\mathbf{K} = (-\frac{4\pi}{3\sqrt{3}}, 0)$ and $\mathbf{K}' = (\frac{4\pi}{3\sqrt{3}}, 0)$), we have $f(\mathbf{K}) = 1$ and $f(\mathbf{K}') = 0$. Furthermore, $\mathbf{K}\cdot\mathbf{a}_1 = -2\pi/3$, $\mathbf{K}\cdot\mathbf{a}_2 = 2\pi/3$, $\mathbf{K}'\cdot\mathbf{a}_1 = 2\pi/3$, and $\mathbf{K}'\cdot\mathbf{a}_2 = -2\pi/3$. The result is that

$$
\hat{V}_{\text{cTBG}}|_{\text{near } \mathbf{K}} \approx W \sum_{\mathbf{k}} \sum_j c_{1,\mathbf{k}-\frac{\mathbf{q}_j}{2}}^\dagger T_j c_{2,\mathbf{k}+\frac{\mathbf{q}_j}{2}} e^{-i\phi_j}, \tag{S15}
$$

with

$$
\begin{aligned}
T_1 &= \sigma_x, \\
T_2 &= e^{2\pi i/3}\sigma_+ + e^{-2\pi i/3}\sigma_-, \\
T_3 &= e^{-2\pi i/3}\sigma_+ + e^{2\pi i/3}\sigma_-.
\end{aligned} \tag{S16}
$$

This exactly describes the continuum model as described in Refs.[S2,S4] with AA tunnelling set to zero (the chiral limit). If we also look at $\mathbf{K}'$ we find that we obtain the same low-energy model up to a unitary transformation.

Importantly, this model has $C_3$ symmetry in addition to $C_2$ symmetry and $T$ (time reversal). These symmetries are contingent on $e^{i\phi_1 + i\phi_2 + i\phi_3} = 1$ (these phases pick an origin of rotation for the symmetry being applied), but are otherwise built into the model.

To test our model against the continuum model, we now compare the density of states. In the continuum model, there is one parameter that controls the physics $\alpha = W/(vk_\theta)$ where $W$ is the tunneling strength between layers, $v = \frac{3}{2}t$ is the Fermi velocity. Therefore, the continuum model has the same physics at $\theta = 1.05°$ as for larger angles such as $\theta = 8.958°$ as we are considering. At larger angles, the continuum model as an approximation for twisted bilayer graphene breaks down due to effects such as band curvature, and the same is true of the lattice model in this section, but we can observe density of state comparisons to see how the well the lattice model is capturing the continuum model as a low-energy approximation. The results are in Fig. S1 and we see that the miniband is captured quite accurately.

In order to simulate this system on finite sizes with periodic boundary conditions, we find that we need $k_\theta = \frac{4\pi}{3} \frac{n}{L}$ for integer $n$ and system size $L$ (for $L^2$ Bravais lattice sites). Then, in order to fully capture an irrational number, such as $k_\theta = \frac{4\pi}{3} \frac{1}{\varphi^5}$ for the golden ratio $\varphi$, we can use the continued fraction expansion defined such that

$$
[a_0; a_1, a_2, \ldots] = a_0 + \cfrac{1}{a_1 + \cfrac{1}{a_2 + \cdots}}, \tag{S17}
$$

if we truncate this such that for instance, $[a_0; a_1, a_2, m]$ with $0 < m \le a_3$, we get a rational that approximates the irrational number ($m = a_3$ being quite a good approximation). Writing $[a_0; a_1, a_2, m] = \frac{n}{L}$ for integers $n, L$ gives us a system size $L$ at which to simulate our system. For example, $1/\varphi^5 = [0; 11, 11, 11, \ldots]$ and we have $[0; 11, 11] = \frac{11}{122}$ while $[0; 11, 11, 3] = \frac{34}{377}$ (approximates of our irrational number for system sizes $L = 122$ and $L = 377$ respectively).

## II. STRUCTURE OF THE SPECTRUM AND SCALING OF THE DENSITY OF STATES

In this section we discuss the numerical method used in the analysis of the spectrum and the finite size effects of the method. We use the kernel polynomial method (KPM) to calculate the density of states $\rho(E)$, which expands $\rho(E)$

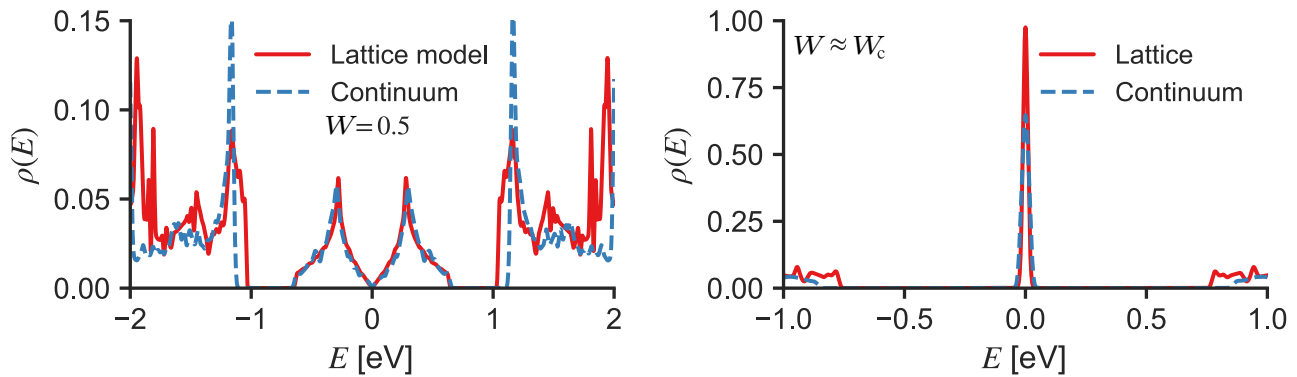

FIG. S1: (left) Comparison of density of states $\rho(E)$ of the lattice and continuum models at $W = 0.5\,\text{eV}$ at an angle $\theta \approx 8.958°$ approximated by taking the fraction $k_\theta = \frac{4\pi}{3}\frac{11}{122}$, we see excellent agreement in the miniband itself, and expected disagreement for higher energy bands beginning with the gap from miniband. (right) Comparison of density of states near the critical value of $W$ for the angle $\theta \approx 8.958°$. For the lattice model we observe this to happen for $W_c \approx 0.875\,\text{eV}$ while for the continuum model we know this occurs at $0.930\,\text{eV}$. The lattice model and the continuum model do not match at high energy, which results in a modified value of the critical value of $W$. We see that the band is quite flat in either case, and just as in the $W = 0.5\,\text{eV}$ case, the higher bands are at slightly different energies.

in terms of Chebyshev polynomials up to an order $N_c$, and we use the Jackson kernel to filter out Gibbs oscillations due to the finite expansion order. To determine the velocity $v$, in two-dimensions for example, we fit the low-$|E|$ asymptote $\rho(E) \approx \rho'(E = 0)|E|$ to extract $\rho'(E = 0) \sim 1/v^2$. Note that in 2D formally $\rho'(E = 0)$ is not just a single derivate due to the $|E|$ scaling, but we use this notation to unify 2D and 3D; the latter it is simply a second derivative. For details on the KPM technique see Ref. S5. We use twisted boundary conditions and we average over random twists to reduce finite size effects. Now we discuss the effect of finite lattice size $L$ and finite cutoff $N_c$ on $\rho(E = 0)$ and $\rho'(E = 0)$.

As an exemplary case we present results here for the "perfect" SOC and the cTBG models defined in the main text. Results on the other models are similar and we also present results on the 1D model below. Figure S2 for the 2D SOC illustrates the dependence on $L$ and $N_c$. For smaller $N_c$ such as $N_c = 2^{12} = 4096$, $\rho'(0)$ for all choice of $L \geq 55$ almost overlap for $W \leq 0.515$. For $N_c = 2^{14} = 16384$, the $\rho'(0)$ data converges as a function of $L$ only for $L \geq 144$. Still, the $L$ convergence is only valid for $W \leq 0.515$. This demonstrates that the observed convergence in $L$ is strongly dependent on $N_C$ and therefore requires studying the scaling in $N_C$ for fixed $L$. In Fig. S3(bottom row) we summarize similar features for cTBG showing Dirac nodes before and after the "magic-angle" and within the metallic phase.

When fixing $L$ and varying $N_c$, the semimetal-to-metal transition becomes sharper as $\rho(E = 0)$ rises more abruptly approaching a sharp step as shown in Fig. S2. This sharpening allows us to pinpoint the location of the transitions accurately, in this case we find $W_c = 0.525 \pm 0.005$ and $W'_c = 0.551 \pm 0.005$. Importantly, the peak of $\rho(0)$ does not shrink as we vary $L$ or $N_c$, providing strong evidence of the presence of the intermediate metallic phase. In addition, we find that $\rho'(0)$ does not saturate as we increase the expansion order, indicating within our numerical accuracy that at the transition $\rho'(0)$ diverges, similar to what was found in 3D[S6]. From the above data of $\rho'(0)$ we determine the scaling exponent $\beta$ defined by $\rho'(0) \sim |W_c - W|^{-\beta}$. We use $\rho'(0)$ data obtained for $N_c = 2^{14}$ and $L = 144$ and we extract $\beta$ from a log-log fit of $1/\rho'(0)$ versus $|W - W_c|$, see Fig. S4.

## A. Dispersion and velocity

In this section we demonstrate the identification of the kinetic velocity as obtained from the twist dispersion with the parameter entering the low-energy asymptote of the DOS. We also compare these numerical results with perturbation theory. We implement twisted boundary conditions by including a factor $e^{i\boldsymbol{\theta}\cdot\mathbf{r}/L}$ for each real space field located at $\mathbf{r}$ and twist vector $\boldsymbol{\theta}$. Each component of $\theta$ takes value in $(0, 2\pi)$ and we compute the energy eigenstates $E(\theta)$ using exact diagonalization for various values of the twist $\theta$. Such a change of boundary condition has no effect on the bulk physics, but effectively moves the origin of the finite size induced momentum grid, so that plotting the spectrum as a function of the twist shows a projection of the dispersion onto $1/L$th of the Brillouin zone. Figure S5 shows the twist dispersion for various models in one, two, and three dimensions, which clearly demonstrates the dramatic flattening

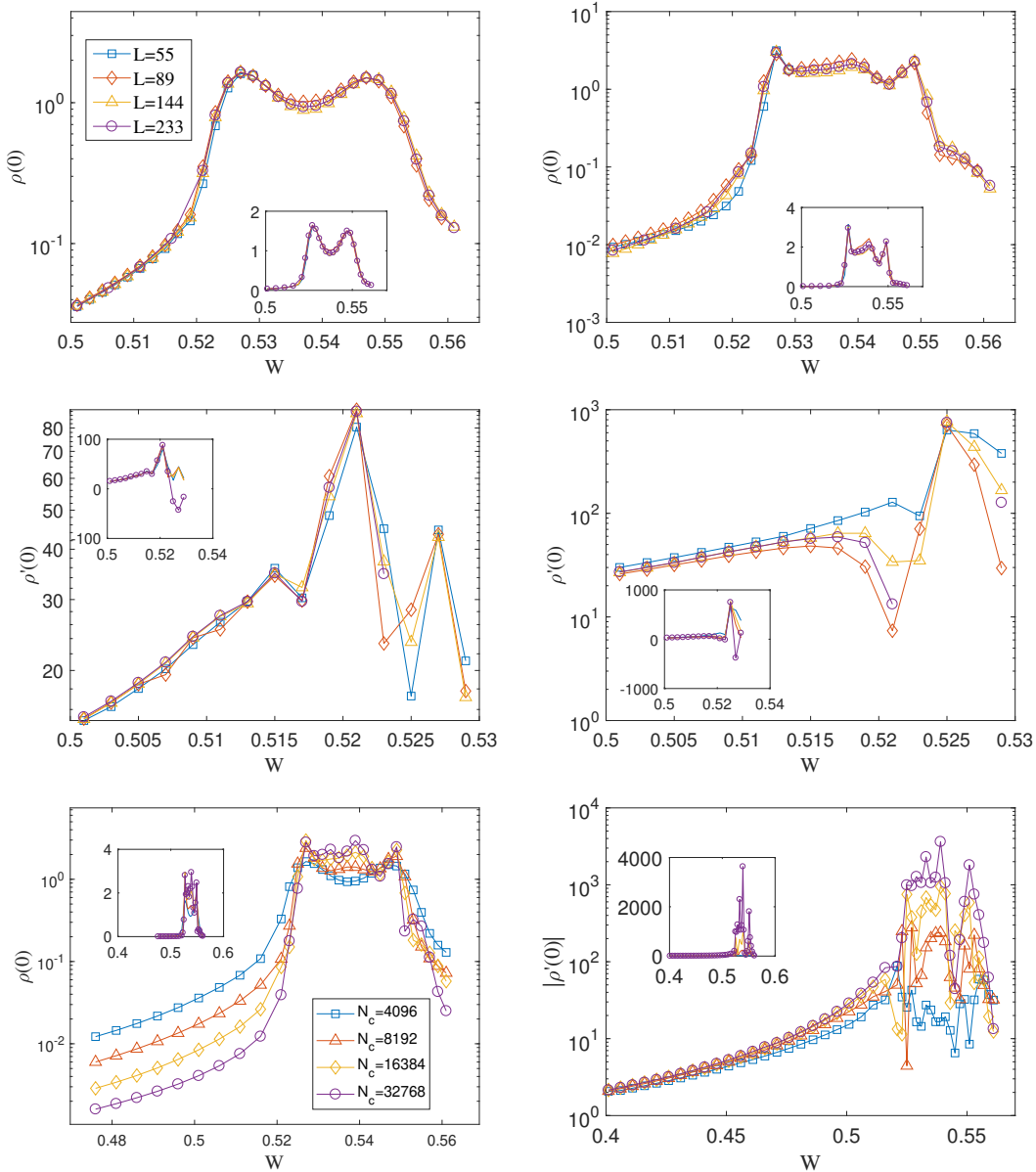

FIG. S2: The DOS $\rho(E=0)$ and its derivative $\rho'(E=0)$ for 2D perfect SOC with $Q = 2\pi F_{n-2}/F_n$ at various $L$ and $N_c$ near the semimetal-to-metal transition $W_c = 0.525 \pm 0.005$ and back to the reentrant semimetal $W'_c = 0.551 \pm 0.005$. Top and middle left: $N_c = 4096$ varying $L$. Top and middle right: $N_c = 16384$ varying $L$. The key is shared across the top four figures. Bottom: $L = 233$, varying $N_c$, with a shared key across the two. The insets are the same plots with linear scale.

of the bands at the transition. These results where obtained for system size $L = 233$ in 1D, $L = 144$ in 2D, and $L = 21$ in 3D. Using the twist dispersion we can estimate the velocity by fitting the lowest energy band near 0 twist to a straight line. We compare the velocity as calculated from the twist dispersion with the KPM result of the DOS and fourth order perturbation theory in Fig. S6, which all agree well.

### B. 1D powerlaw hopping model

The parameter $\sigma$ defined in equation (S6) determines the behavior of the dispersion relation near $k = 0$. This can be seen directly from the twist dispersion in Fig. S5. We present detailed results for the 1D LRH model in Fig. S7.

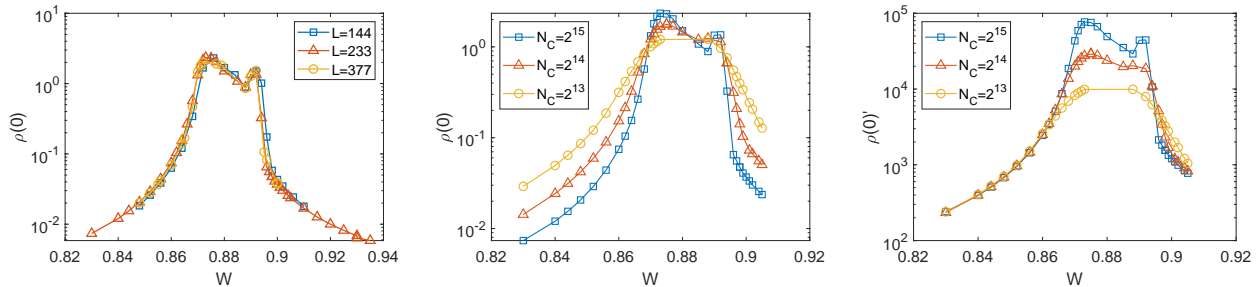

FIG. S3: The DOS (left and middle) $\rho(E = 0)$ and its derivative (right) $\rho'(E = 0)$ for cTBG at $k_\theta = 2\pi F_{n-5}/F_n$ for different system sizes (left) and for different KPM expansion orders for $L = F_n = 233$ at various $N_c$ (middle and right) near the semimetal-to-metal transition $W_c = 0.8725 \pm 0.001$ and back to the reentrant semimetal $W_c' = 0.892 \pm 0.002$.

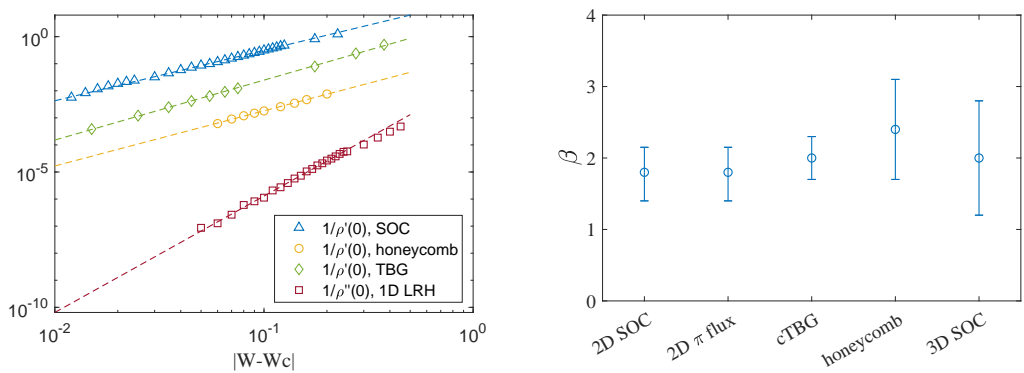

FIG. S4: Estimate of the scaling exponent $\beta$. Left: Extracting $\beta$ from fitting (dashed lines) $1/\rho'(0)$ in 2D and $1/\rho''(0)$ in 1D for $\sigma = 1/3$ versus $|W - W_c|$ on a log-log plot displaying a clear power law. For clarity we have shifted the data of the honeycomb (HC) model so that it doesn't overlap with the cTBG model. Right: $\beta$ estimate for the models we investigated in this paper, note that the estimate of $\beta$ for the 3D Weyl model quotes the result from Ref.[S6] and for the 1D long range hopping model we find $\beta = 4 \pm 0.8$ for $\sigma = 1/3$.

The DOS depends on $\sigma$ by $\rho(E) \sim |E|^{1/\sigma - 1}$ (Ref. S7), which is demonstrated in Fig. S7. In the following we present detailed results for $\sigma = 1/3$ and leave the full exploration of this 1D model for future work. Focusing on $\sigma = 1/3$ is numerically advantageous since as we approach the transition the scaling $\rho(E) \sim |W - W_c|^{-\beta}|E|^2$ allows us to use the second derivative of the DOS $\rho''(0)$ to estimate $\beta$ and we can compute $\rho''(0)$ accurately using the KPM. Notice that the power-law remains constant when varying $W$ in the semimetal phase, showing the 1D model is also stable to a weak quasiperiodic potential. Upon approaching the transition we find $\rho''(0)$ displays a clear divergence with no sign of saturation as we increase the expansion order (see Fig. S7), similar to the 2D model we have discussed above. We find $W_c = 2.05 \pm 0.03$ and from the power-law scaling $\rho''(0) \sim |W - W_c|^{-\beta}$ we extract $\beta = 4.0 \pm 0.8$ for $\sigma = 1/3$, see Fig. S4. Distinct from our results in 2D and 3D the transition in the 1D model is accompanied by real space localization. To demonstrate this we calculate the IPR in real space and momentum space at zero energy. The real space IPR becomes finite and momentum space IPR vanishes near the critical $W_c$. In addition, when the momentum space IPR goes to zero the DOS becomes non-zero demonstrating that the generation of DOS is tied with momentum space delocalization, similar to the higher dimensional models.

# III. ANALYTIC RESULTS

This part of the supplement is devoted to the summary of details on analytical arguments presented in the main text.

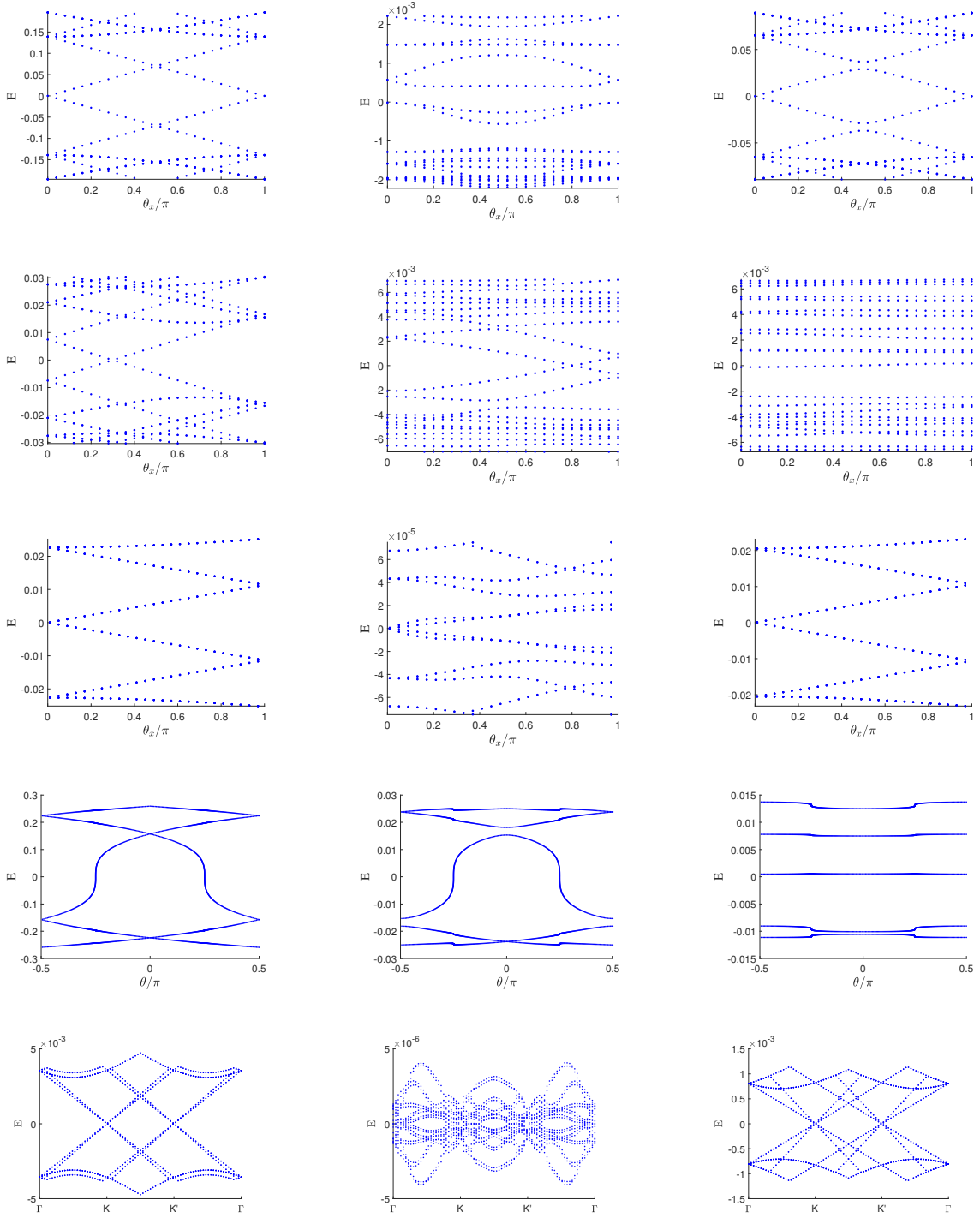

FIG. S5: Twist dispersion for various models displaying the characteristic magic angle feature of flat bands. All results shown except for the second and last rows (honeycomb model and cTBG respectively) are on the twist trajectory $\theta_y = 0$, and for 3D $\theta_z = 0$. Whereas for the honeycomb model and cTBG, $\theta_y = -\theta_x$ (going through high symmetry points as displayed in the last row). First row: 3D SOC model investigated in Ref. S6 with $Q = 2\pi F_{n-2}/L$, $L = 21$, $W = 0.1, 0.384, 0.5$ corresponding to before the transition, in the metallic phase and after the transition. Second row: 2D honeycomb model at $Q = 2\pi F_{n-3}/L$, $L = 144$, $W = 0.8, 1.0, 1.08$ representing the states well before the transition, right before the transition, and a flattened band. Third row: 2D SOC model at $Q = 2\pi F_{n-2}/L$, $L = 144$, $W = 0.35, 0.54, 0.8$ depicting states before the transition, in the metallic phase and after the transition. We have numerically checked that the twist dispersion of the $\pi$-flux model coincides with the 2D SOC, see the arguments exposed in Fig. S8, below. Fourth row: 1D power law hopping model at $Q = 2\pi F_{n-3}/L$, $\sigma = 1/3$, $L = 1597$, $W = 0.5, 1.7, 2.051$ of states well before the transition, the formation of a miniband, and a flattened band. Last row: 2D cTBG model at $k_\theta = 4\pi/3 \times 21/L$, $L = 233$, $W = 0.8, 0.876, 0.9$ for states before the magic-angle, at the magic-angle, and after the magic-angle respectively.

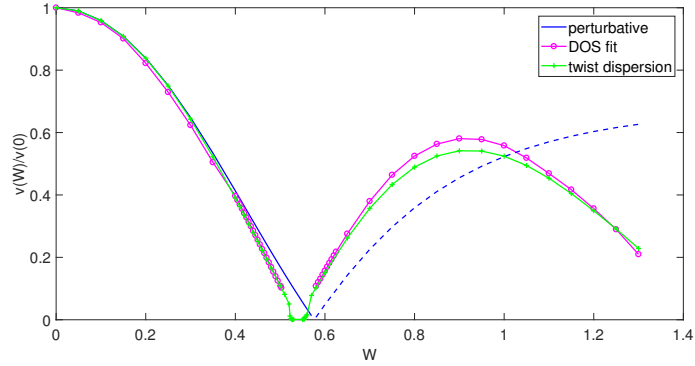

FIG. S6: Velocity for the 2D perfect SOC model at $Q = 2\pi F_{n-2}/F_n$, as obtained from fitting the DOS near $E = 0$ (with $L = 144$, $N_c = 16384$ KPM result), fitting the twist dispersion near 0 twist (with $L = 144$ exact diagonalization), and the perturbative calculation, Eq. (4) of the main text. The dashed line is the sign-reversed perturbative result for post-transition $W$.

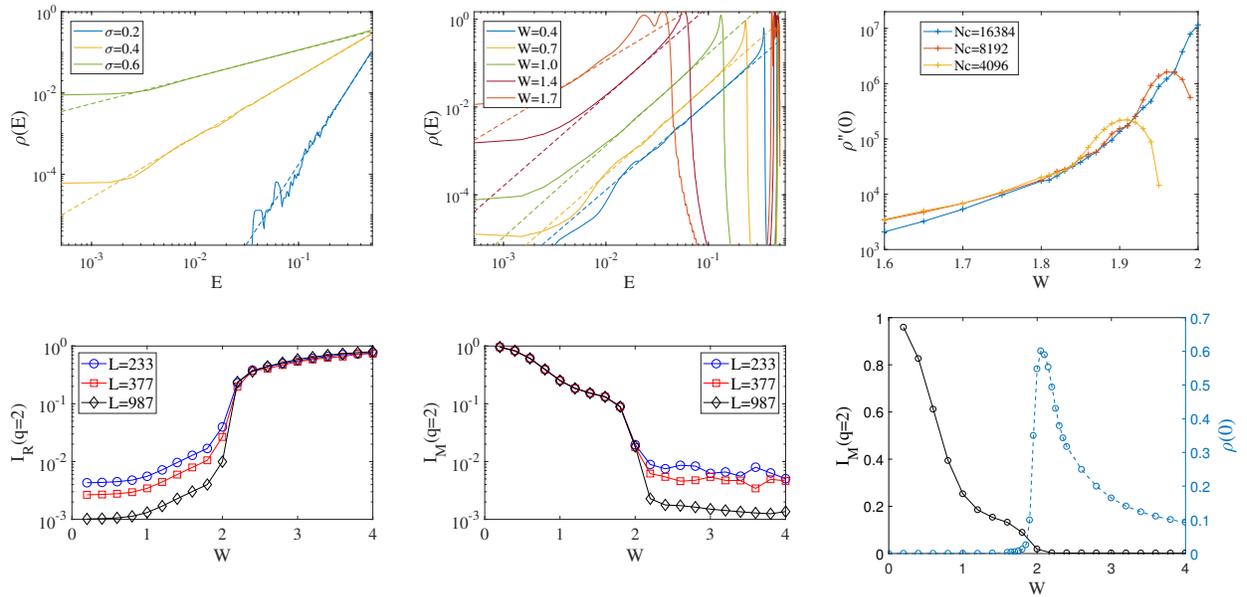

FIG. S7: Results for the long-range hopping model with $Q/2\pi = F_{n-3}/F_n$. Top left: Density of states $\rho(E)$ as a function of $E$ on a log-log scale calculated using KPM at $L = 1597$, with an $N_c = 4096$ with $W = 0$. By varying $\sigma$ the power-law decreases following $\rho(E) \sim |E|^{1/\sigma - 1}$. Top middle: Fixed $\sigma = 1/3$ and varying $W$, when $W$ is small, the power law is well preserved. As the model approaches the transition the power law regime is pushed to lower and lower energy due to the formation of minibands. The solid line represent actual data, and dashed lines are fits to the expected form $\rho(E) \sim |E|^{1/\sigma - 1} = E^2$. To accurately extract the scaling of the prefactor of the DOS, i.e. $\rho(E) \sim |W - W_c|^{-\beta}|E|^{1/\sigma - 1}$, we turn to the second derivative of $\rho(E)$. Top right: The second derivative $\rho''(0)$ as a function of $W$ for various $N_C$ with $L = 28657$ displaying a clear divergence as the transition is approached, signifying the DOS is becoming non-analytic. As shown in Fig. S4, we find $\rho''(0)$ diverges with $|W - W_c|$ in a power law fashion. To avoid the issues involved with fitting, that we find are most severe in this 1D model we use $\rho''(0)$ computed directly with the KPM. Bottom left and middle: Real space and momentum space inverse participation ratio (IPR) of energy eigenstate closest to $E = 0$, denoted as $\mathcal{I}_R(q = 2)$ and $\mathcal{I}_M(q = 2)$ respectively, calculated for various $L$ and averaged over 200 realizations. The $L$ dependence clearly shows that when the real space IPR is delocalized $\mathcal{I}_R(q = 2) \sim |E|^{1/\sigma - 1} = 1/L$ the momentum space IPR is localized and vice-versa. This demonstrates the transition in 1D goes from a semimetal to an Anderson insulator. Bottom right: The momentum space IPR $\mathcal{I}_M(q = 2)$ (left vertical axis) and the zero energy DOS $\rho(0)$ (right vertical axis) for $L = 987$ as a function of $W$, which demonstrates that the zero energy DOS becomes non-zero when the wavefunctions delocalize in momentum space.

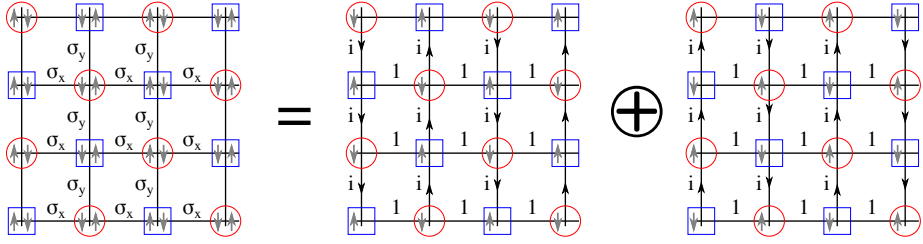

FIG. S8: Graphic demonstration that the model of perfect SOC in 2D is a direct sum of two decoupled $\pi$ flux models. The model of perfect SOC, on the left of the equality sign, is characterized by direction dependent hopping matrices. Using blue squares and red circles to depict the bipartition, hopping only connects $|\square, \uparrow\rangle$ with $|\circ, \downarrow\rangle$, and *separately* $|\square, \downarrow\rangle$ with $|\circ, \uparrow\rangle$. The hopping in $y$-direction is imaginary and directed (this results from the asymmetry of $\sigma_y$) and, in conclusion, leads to the inclusion of a flux $\pi$ per plaquette. The onsite potential does note violate the described block-diagonalization.

## A. Perturbative calculation of velocity renormalization

In this section we present the perturbative calculation of velocity renormalization using the language of retarded Green's functions,

$$\hat{G}_0(E) = [E + i\eta - \hat{T}]^{-1}, \quad \hat{G}(E) = [E + i\eta - \hat{T} - \hat{V}]^{-1}, \tag{S18a}$$

and are interested in diagonal components $G_{\mathbf{k},\mathbf{k}'}$ with $\mathbf{k} = \mathbf{k}'$, only (e.g. for the DOS we only need $\rho(E) = -(1/\pi)\mathrm{Im}\sum_{\mathbf{k}} \mathrm{Tr} G_{\mathbf{k},\mathbf{k}}(E)$). We define the self energy at momentum $\mathbf{k}$ by all diagrams which are $G_0(\mathbf{k}, E)$ irreducible and write

$$G_{\mathbf{k},\mathbf{k}}(E) = [G_0(\mathbf{k}, E)^{-1} - \Sigma(\mathbf{k}, E)]^{-1}. \tag{S18b}$$

We expand about a given node $\mathbf{K}_i$ of the dispersion $T(\mathbf{K}_i + \mathbf{p}) \simeq T(\mathbf{K}_i) + h^{(\mathbf{K}_i)}(\mathbf{p})$ to leading order in $p \ll 1/a$. For models which satisfy the symmetry constraints exposed in the main text (see also Sec. III D) $\Sigma(\mathbf{K}_i + \mathbf{p}, E) = E\Sigma_E + h(\mathbf{p})\Sigma_p$ to leading order in $E, p$. Henceforth, we choose the energy offset such that $T(\mathbf{K}) = 0$. Then,

$$G_{\mathbf{k},\mathbf{k}}(E) = Z[E - (v/v_0)h(\mathbf{k})]^{-1} \quad \text{with } Z^{-1} = 1 - \Sigma_E \text{ and } v/v_0 = (1 + \Sigma_p)Z. \tag{S19}$$

In this section we evaluate the self energy to leading and, for some models, next to leading order in powers of $W$ and summarize them in Tab. I. A discussion of infinite order perturbation theory can be found in Sec. III D.

To illustrate the procedure we analyze the model of 2D perfect SOC for which the states at small $\mathbf{k}$ with Hamiltonian $H(\mathbf{k}) = t(\sin(k_x)\sigma_x + \sin(k_y)\sigma_y) \simeq t\mathbf{k} \cdot \boldsymbol{\sigma}$ are connected to the states at $\mathbf{k} \pm Q\hat{e}_{x,y}$ and therefore to leading order perturbation theory

$$\Sigma^{(2)}(\mathbf{k}) = (W/2)^2 \sum_{\pm} \frac{1}{E - H(\mathbf{k} \pm Q\hat{e}_x)} + x \leftrightarrow y \simeq -E4\alpha^2 - t\mathbf{k} \cdot \boldsymbol{\sigma}(2\alpha^2(1 - \cos(Q)) \tag{S20}$$

For the next to leading order, all states at Manhatten distance $2Q$ from the origin are integrated out and we obtain

$$\Sigma^{(4)}(\mathbf{k}) \simeq -\frac{E}{16}\left(\frac{W}{t}\right)^4 \left(4\cos(Q) + 10\cos(2Q) + 11\right)\csc^4(Q)\sec^2(Q)) \\ + \left(\frac{W}{t}\right)^4 \frac{t\mathbf{k} \cdot \boldsymbol{\sigma}}{16}(4 - 5\cos(Q) + 6\cos(2Q))\csc(Q)^4\sec(Q) \tag{S21}$$

This is the origin of Eq. (4) of the main text. It turns out that the results obtained for the 2D model of perfect SOC directly apply to the $\pi$ flux model. This is best graphically shown, see Fig. S8: the model of 2D perfect SOC is a direct sum of two $\pi$-flux models which in the absence of interactions completely decouple. By consequence, all single particle results obtained for model of 2D perfect SOC also hold for the $\pi$-flux model.

| model | $Z^{-1}$ | $v/v_0$ |
|---|---|---|
| 2D SOC | $1 + \left[\dfrac{W}{t\sin(Q)}\right]^2$ | $Z\left(1 - \dfrac{1-\cos(Q)}{2}\left[\dfrac{W}{t\sin(Q)}\right]^2\right)$ |
| 3D SOC | $1 + \dfrac{3}{2}\left[\dfrac{W}{t\sin(Q)}\right]^2$ | $Z\left(1 - \dfrac{2-\cos(Q)}{2}\left[\dfrac{W}{t\sin(Q)}\right]^2\right)$ |
| 1D LRH | $1 + 2\left[\dfrac{W}{t|\sin(Q)|^\sigma}\right]^2$ | $Z$ |
| 2D cTBG | $1 + \dfrac{W^2}{t^2}\dfrac{\left[\cos\left(\frac{3k_\theta}{2}\right)+6+2\cos\left(\frac{3k_\theta}{4}\right)\right]}{12\sin^2\left(\frac{3k_\theta}{4}\right)}$ | $Z\left(1 - \dfrac{W^2}{t^2\sin^2\left(\frac{3k_\theta}{4}\right)}\dfrac{\left[2\cos\left(\frac{3k_\theta}{2}\right)-16\cos\left(\frac{3k_\theta}{4}\right)+4\cos\left(\frac{9k_\theta}{4}\right)+5\cos(3k_\theta)-49\right]}{72}\right)$ |
| HC, $\boldsymbol{\delta}_\mu = \mathbf{d}_\mu$ | $1 + \dfrac{W^2}{t^2}\dfrac{3}{8\sin^2(3Q/4)}$ | $Z\left(1 - \dfrac{W^2}{t^2}\dfrac{1/4}{1+2\cos(Q/2)}\right)$ |
| HC, $\boldsymbol{\delta}_\mu = O_{\pi/2}\mathbf{d}_\mu$ | $1 + \dfrac{W^2}{t^2}\dfrac{3[2+\cos(\sqrt{3}Q/2)]}{8\sin^2(3\sqrt{3}Q/4)}$ | $Z\left(1 + \dfrac{W^2}{t^2}\dfrac{5+4\cos(\sqrt{3}Q/2)}{4[1+\cos(\sqrt{3}Q/2)]^2}\right)$ |
| HC, $\boldsymbol{\delta}_\mu = \mathbf{d}_{1,2}$ | $1 + \dfrac{W^2}{t^2}\dfrac{1}{4\sin^2(3Q/4)}$ | $Z\left(1 - \dfrac{W^2}{t^2}\dfrac{\begin{matrix}3[\cos(Q/2)-2\cos(Q)] & -\sqrt{3}[\cos(Q/2)+2\cos(Q)]\\ -\sqrt{3}[\cos(Q/2)+2\cos(Q)] & 5\cos(Q/2)-2\cos(Q)\end{matrix}}{48\sin(3Q/4)^2}\right)$ |

TABLE I: Perturbative corrections to quasiparticle weight $Z$ and velocity $v/v_0$ for a variety of magic angle models. Note that for the honeycomb model and $\boldsymbol{\delta}_\mu = \mathbf{d}_{1,2}$, the symmetry protection of nodes is lost. It implies a relocation of $K-$point node $\delta\mathbf{k} = W^2(1, -\sqrt{3})^T/[12t^2(1 + 2\cos(Q/2))]$ and a distorted velocity matrix. For cTBG, the momentum dependent self-energy has the form $\Sigma_\mathbf{p}h(p_x, p_y) + \tilde\Sigma_\mathbf{p}h(p_y, -p_x)$ and we extract $v/v_0 = Z(1 + \Sigma_\mathbf{p})$.

### B. Renormalization of interactions

In this section we present an analytical estimate of the renormalization of the interaction upon projection onto certain minibands and approaching the transition for the model of 2D SOC. Let the bare ($W = 0$) model in the continuum be written as ($\mathbf{K}_i$ are various Dirac/Weyl nodes, with linear $\mathbf{k}\cdot\mathbf{p}$ Hamiltonian $h^{(\mathbf{K}_i)}(\mathbf{p})$)

$$S = \sum_{\mathbf{K}_i}\int^{1/a}(dp)\int d\tau \bar{c}^{(\mathbf{K}_i)}(\mathbf{p})[\partial_\tau + h^{(\mathbf{K}_i)}(\mathbf{p})]c^{(\mathbf{K}_i)}(\mathbf{p})$$
$$+ \sum_{\mathbf{K}_{1,2,3,4}}\int^{1/a}\prod_{j=1}^4(dp_j)\int d\tau(2\pi)^d g_{\{\mathbf{K}_i\}}\delta_{\mathbf{K}_1+\mathbf{K}_3-\mathbf{K}_2-\mathbf{K}_4}\delta(\mathbf{p}_1+\mathbf{p}_3-\mathbf{p}_2-\mathbf{p}_4)\frac{[\bar{c}^{(\mathbf{K}_1)}(\mathbf{p}_1)c^{(\mathbf{K}_2)}(\mathbf{p}_2)][\bar{c}^{(\mathbf{K}_3)}(\mathbf{p}_3)c^{(\mathbf{K}_4)}(\mathbf{p}_4)]}{|\mathbf{p}_1-\mathbf{p}_2|^\sigma}.$$
$$(S22)$$

The spectrum of $h^{(\mathbf{K}_i)}(\mathbf{p})$ has the form $v_0|\mathbf{p}|$ with bare value $v_0 \sim ta$ and, for contact interaction ($\sigma = 0$), $g_{\{\mathbf{K}_i\}} \sim Ua^d$, while for Coulomb interaction ($\sigma = d - 1$) $g_{\{\mathbf{K}_i\}} \propto \delta_{\mathbf{K}_1,\mathbf{K}_2}$. Perturbation theory indicates a dimensionless parameter

$$\frac{g_{\{\mathbf{K}_i\}}}{v_0/a^{d-1}} \sim \frac{U}{t}\ \text{for } \sigma = 0\ \text{(contact interaction)},\qquad (S23)$$

$$\alpha = \frac{g_{\{\mathbf{K}_i\}}}{v_0}\ \text{for } \sigma = d - 1\ \text{(Coulomb interaction)}.\qquad (S24)$$

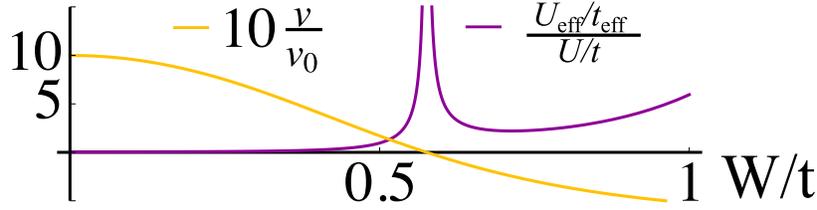

FIG. S9: Divergence of contact interaction according to Eq. (S27) for the model of 2D SOC. Here, the fourth order perturbative self energy was employed and we used $\gamma = 1/5$.

We now consider the effect of integrating out high energy states and projecting onto a miniband with effective Brillouin zone size $1/a'$. This leads to

$$
\begin{aligned}
S = & \sum_{\mathbf{K}_i} \int^{1/a'} (dp') \int d\tau \bar{c}_<^{(\mathbf{K}_i)}(\mathbf{p}') Z^{-1} [\partial_\tau + \frac{v}{v_0} h^{(\mathbf{K}_i)}(\mathbf{p}')] c_<^{(\mathbf{K}_i)}(\mathbf{p}') \\
& + \sum_{\mathbf{K}_{1,2,3,4}} \int^{1/a'} \prod_{j=1}^4 (dp_j') \int d\tau (2\pi)^d g_{\{\mathbf{K}_i\}} \delta_{\mathbf{K}_1 + \mathbf{K}_3 - \mathbf{K}_2 - \mathbf{K}_4} \delta(\mathbf{p}_1' + \mathbf{p}_3' - \mathbf{p}_2' - \mathbf{p}_4') \frac{[\bar{c}_<^{(\mathbf{K}_1)}(\mathbf{p}_1') c_<^{(\mathbf{K}_2)}(\mathbf{p}_2')][\bar{c}_<^{(\mathbf{K}_3)}(\mathbf{p}_3') c_<^{(\mathbf{K}_4)}(\mathbf{p}_4')]}{|\mathbf{p}_1' - \mathbf{p}_2'|^\sigma}.
\end{aligned}
\tag{S25}
$$

The renormalizations $Z$ and $v/v_0$ originate from scalar and matrix components of the self-energy and were calculated perturbatively above. We now first rescale $p' = \frac{a}{a'}p$ with $p \in (0, 1/a)$ and then define $c_<^{(\mathbf{K})}(ap'/a') Z^{-1/2} (a/a')^{d/2} = c^{(\mathbf{K})}(\mathbf{p})$. Under this rescaling, we restore the form of Eq. (S22), including its UV cut-off $1/a$, but obtain the rescaling $v_0 \to v a/a'$, $g \to g(a/a')^{d-\sigma} Z^2$. From this we obtain the final formula for renormalization of the dimensionless coupling constant

$$
\frac{U_{\text{eff}}/t_{\text{eff}}}{U/t} = \gamma \left(\frac{a}{a'}\right)^{d-1} \frac{Z^2}{v/v_0} \qquad \text{(contact interaction)},
\tag{S26}
$$

$$
\frac{\alpha_{\text{ren.}}}{\alpha_{\text{bare}}} = \frac{Z^2}{v/v_0} \qquad \text{(Coulomb)}.
\tag{S27}
$$

Here, $\gamma$ is an unknown constant of order unity which depends on details of the cut-off of the linearized theory. Importantly, the integration reduces the bare contact interaction by a factor $(a/a')^{d-1}$, except in the closest vicinity of the magic angle where the vanishing velocity overtakes the reduction, see Fig. S9.

### C. Relationship to number theory

In this section we show the relationship of the sequence of relevant perturbative processes with certain well known sequences from number theory. Starting from the scattering process of order $l_1 = 1$ we want to determine the sequence $\{l_n\}_{n=1}^\infty$ for which the $l_n$th order momentum transfer carves out smaller minibands than the $l_{n-1}$th order. In formulae, this implies for the 2D model of perfect SOC of the main text and arbitrary incommensuration wavevector $Q$ the condition $\sin^2(l_n Q) < \sin^2(l_{n-1} Q)$. We now concentrate on the specific case $Q = 2\pi/\phi^2 = \pi(3 - \sqrt{5})$. For this situation, the defining condition on the sequence of $l_n$ is $\sin^2(\pi l_n \sqrt{5}) < \sin^2(\pi l_{n-1} \sqrt{5})$. The sequence $\{l_n\}_{n=1}^\infty$ for which $l_n \sqrt{5}$ successively approaches integers is the sequence of denominators of the leading rational approximants, i.e. the sequence of denominators of continued fraction convergents of $\sqrt{5}$ (OEIS ID A001076). This sequence is also half the value of the even fibonaccis $l_n = F_{3n}/2$.

This sequence also connects to the formation of minibands as found with the finite size $Q_n = 2\pi F_{n-2}/F_n$. Intuitively, when $F_n$ is even ($n$ is a multiple of 3), then at order $F_n/2$, the Dirac nodes gap out, but then for $F_{n+1}$ and $F_{n+2}$ this perturbative gap must have moved to small but finite energy, forming the miniband. This motivates using $Q_{3n+1}$ and $Q_{3n+2}$ to study the effective model of successive minibands, as we do in Sec. VI.

Therefore, we offer a proof that connects this sequence to the perturbative minibands which requires the following facts about Fibonaccis

- $F_{3m}$ is even while $F_{3m+1}$ and $F_{3m+2}$ is odd.

- Catalan's Identity: $(-1)^{n-r} F_r^2 = F_n^2 - F_{n+r} F_{n-r}$.

- $(-1)^n = F_n F_{n-1} - F_{n-2} F_{n+1}$.

To determine the order of perturbation theory where a gap is opened, we need to find an integer $g_n$ such that $g_n Q_n = 2\pi \times q/F_n$ such that $q$ is an integer closest to $F_n/2$ modulo $F_n$. This is accomplished by the following theorem

**Theorem.** *Let $n = 3m + r$ for $r = 0, 1, 2$, then the integer $g_n = F_{3m}/2$ is the smallest integer such that $g_n F_{n-2} \equiv \frac{F_n + \delta_n}{2} \mod F_n$ with integer $|\delta_{3m}| \leq 1$. In particular, $\delta_{3m} = 0$, $\delta_{3m+1} = (-1)^m$, and $\delta_{3m+2} = (-1)^{m+1}$.*

*Proof.* We break this up into cases. Since $F_{n-2}$ does not divide $F_n$, we merely need to find a $g_n$ that satisifies the relevant cases in order to find the unique $g_n$, as long as $g_n < F_n$.

**Case 1:** $r = 0$. For this case we can prove the above by noting that if $g_{3m} = F_{3m}/2$, and $F_{n-2}$ is necessarily odd, then (We use $N$ to represent an arbitrary, unimportant, integer.)

$$g_n F_{n-2} = \tfrac{1}{2} F_{3m}(2N+1) \equiv \tfrac{1}{2} F_{3m} \mod F_n, \tag{S28}$$

with $\delta_{3m} = 0$.

The value $g_n$ is smallest since $\delta_n$ must vary by 2 in order for the equation $\frac{F_n + \delta_n}{2}$ to remain integer valued, and the condition $|\delta_n| < 1$ prevents that.

**Case 2:** $r = 1$. For this case we observe using the previous fact about Fibonaccis that

$$F_{3m} F_{3m-1} = F_{3m+1} F_{3m-2} + (-1)^{3m}, \tag{S29}$$

and therefore since $F_{3m-2}$ is odd

$$\tfrac{1}{2} F_{3m} F_{3m-1} = (N + \tfrac{1}{2}) F_{3m+1} + \tfrac{1}{2}(-1)^{3m} \equiv \frac{F_{3m+1} + (-1)^{3m}}{2} \mod F_n, \tag{S30}$$

and therefore $g_{3m+1} = F_{3m}/2$ with $\delta_{3m+1} = (-1)^m$.

In order to show that $g_n$ is smallest, only $\delta_{3m+1} = -(-1)^m$ could be a problem, but we notice that $\delta_{3m+1} = -(-1)^m$ is satisfied (uniquely) for $g_n = F_{3m+1} - F_{3m}/2 > F_{3m}/2$.

**Case 3:** $r = 2$. For this case we take Catalan's identity with $r = 2$ to get

$$F_{3m}^2 = F_{3m+2} F_{3m-2} - (-1)^{3m}, \tag{S31}$$

and therefore since $F_{3m-2}$ is odd

$$\tfrac{1}{2} F_{3m} F_{3m} = (N + \tfrac{1}{2}) F_{3m+2} - \tfrac{1}{2}(-1)^{3m} \equiv \frac{F_{3m+2} - (-1)^{3m}}{2} \mod F_n, \tag{S32}$$

so again $g_{3m+2} = F_{3m}/2$ with $\delta_{3m+2} = (-1)^{m+1}$.

In order to show that $g_n$ is smallest, only $\delta_{3m+1} = (-1)^m$ could be a problem, but we notice that $\delta_{3m+1} = (-1)^m$ is satisfied (uniquely) for $g_n = F_{3m+2} - F_{3m}/2 > F_{3m}/2$. $\square$

Therefore, the order of perturbation theory that opens up a gap nearest to $E = 0$ for $Q = 2\pi F_{3m+r-2}/F_{3m+r}$ for $r = 0, 1, 2$ is $F_{3m}/2$.

### D. Generality of the magic-angle phenomenon - symmetry protection

In this part of the supplement we discuss the generality of our findings by highlighting the general condition for the appearance of the magic angle phenomenon, namely the stability of the semimetal at weak coupling.

We concentrate on nodes in the kinetic term $\hat{T}$ which are protected by a symmetry group $G_T$. For example, this analysis applies to each model we have considered in 2D as well as Dirac semimetals in 3D. Note that in general $G_T$ is a subgroup of all symmetry operations of the kinetic term. Let $U_{S_T}$ be the representation of $S_T \in G_T$ in the (e.g. spinorial) Hilbert space, then the symmetry of the Hamiltonian implies $T(\mathbf{k}) = U_{S_T}^\dagger T(S_T \mathbf{k}) U_{S_T}$. We concentrate on high symmetry points where $S_T \mathbf{K} = \mathbf{K}$, $\forall S_T \in G_T$. Then, a non-trivial representation implies degeneracy in view of $[T(\mathbf{K}), U_{S_T}] = 0$ $\forall S_T \in G_T$ (formally, two non-commuting $U_{S_T}$ are needed). We further assume a group $G_V$ of spatial (point group) symmetries of the quasiperiodic background $\tilde{V}$, such that

$$V(\mathbf{x}) = \sum_{S_V \in G_V} \tilde{U}_{S_V} W \tilde{U}_{S_V}^\dagger e^{iQ\mathbf{x} \cdot S_V \hat{e}_0} + h.c.. \tag{S33}$$

Here now, $\tilde{U}_{S_V}$ is the representation of $S_V \in G_V$ and $\hat{e}_0$ is an arbitrary vector in $\mathbb{R}^d$.

In this section, we consider Eq. (S18) formally to all orders in perturbation theory. The semimetallic behavior persists if a) $\Sigma(\mathbf{k})$ is hermitian and b) $T(\mathbf{k}) + \Sigma(\mathbf{k})$ has the same symmetry protected touching point as $T(\mathbf{k})$, i.e. if $\Sigma(\mathbf{k})$ respects the symmetries ensuring the semimetal. In view of the incommensuration, perfect resonance is formally absent to any order in perturbation theory and therefore, the decay rate $1/\tau \sim \sum_{\mathbf{k}'} |\mathbf{T}_{\mathbf{k},\mathbf{k}'}|^2 \delta(E_{\mathbf{k}} - E_{\mathbf{k}'})$ (more generally: the anti-hermitian part of the self-energy) vanishes ($\mathbf{T}_{\mathbf{k},\mathbf{k}'}$ denotes the T-matrix). Thus $a$) is fulfilled and $1/\tau \neq 0$ signals the breakdown of perturbation theory (spontaneous unitarity breaking). We can then show to all orders in perturbation theory that the semimetal is stable provided $G_T$ is a subgroup of $G_V$.

We proceed to the proof of $\Sigma(\mathbf{k}) = U_{S_T}^\dagger \Sigma(S_T \mathbf{k}) U_{S_T}$ under the outlined assumptions. To get a feeling, we first consider second order perturbation theory.

$$\Sigma^{(2)}(\mathbf{k}) = \sum_{S_V \in G_V} \tilde{U}_{S_V} W \tilde{U}_{S_V}^\dagger [E^+ - T(\mathbf{k} + QS_V \hat{e}_0)]^{-1} \tilde{U}_{S_V} W^\dagger \tilde{U}_{S_V}^\dagger. \tag{S34}$$

We compare to

$$U_{S_T}^\dagger \Sigma^{(2)}(S_T \mathbf{k}) U_{S_T} = \sum_{S_V \in G_V} U_{S_T}^\dagger \tilde{U}_{S_V} W \tilde{U}_{S_V}^\dagger U_{S_T} [E^+ - U_{S_T}^\dagger T(S_T \mathbf{k} + QS_V \hat{e}_0) U_{S_T}]^{-1} U_{S_T}^\dagger \tilde{U}_{S_V} W \tilde{U}_{S_V}^\dagger U_{S_T}$$

$$= \sum_{S_V \in G_V} U_{S_T}^\dagger \tilde{U}_{S_V} W \tilde{U}_{S_V}^\dagger U_{S_T} [E^+ - T(\mathbf{k} + QS_V^{-1} S_V \hat{e}_0)]^{-1} U_{S_T}^\dagger \tilde{U}_{S_V} W^\dagger \tilde{U}_{S_V}^\dagger U_{S_T} \tag{S35}$$

This expression is invariant provided the action of $S_T$ onto $G_V$ is a bijection of $G_V$ onto itself $\forall S_T \in G_T$, i.e. $S_T S_V \in G_V \; \forall S_V \in G_V$ and $S_T G_V = G_V$ as this allows to uniquely relabel the summation index. Taking $S_V = \mathbf{1}$ implies that $S_T \in G_V$ and hence $G_T$ is a subgroup of $G_V$. By consequence, the representation in the Hilbert space fulfills $\tilde{U}_{S_T^{-1} S_V} = U_{S_T}^\dagger \tilde{U}_{S_V}$ and $\Sigma^{(2)}(\mathbf{k})$ is invariant under the symmetries protecting the semimetal. We now continue with the next order $\Sigma^{(4)}$, from there the generality of the statement becomes apparent,

$$\Sigma^{(4)}(\mathbf{k}) = \sum_{S_V \in G_V} \sum_{\substack{S_V' \in G_V \\ S_V' \hat{e}_0 \neq -S_V \hat{e}_0}} \tilde{U}_{S_V} W \tilde{U}_{S_V}^\dagger [E^+ - T(\mathbf{k} + QS_V \hat{e}_0)]^{-1}$$

$$\tilde{U}_{S_V'} W \tilde{U}_{S_V'}^\dagger [E^+ - T(\mathbf{k} + QS_V \hat{e}_0 + QS_V' \hat{e}_0)]^{-1}$$

$$\tilde{U}_{S_V'} W^\dagger \tilde{U}_{S_V'}^\dagger [E^+ - T(\mathbf{k} + QS_V \hat{e}_0)]^{-1} \tilde{U}_{S_V} W^\dagger \tilde{U}_{S_V}^\dagger. \tag{S36}$$

The exclusion $S_V' \hat{e}_0 + S_V \hat{e}_0 \neq 0$ ensures the irreducibility with respect to $G_{\mathbf{k}}$. Again we can apply an $S_T$ transformation and exploit the two conditions exposed above to relabel both $S_V$ and $S_V'$. This implies the invariance of $\Sigma_{(4)}$. This procedure can be used to arbitrary order in perturbation theory.

## IV. WAVEPACKET DYNAMICS

### A. Non-interacting Wavepacket Expansion as a Probe of the Single Particle Quantum Phase Transition

In this section we demonstrate that wavepacket dynamics can be used as a tool to observe the single particle quantum phase transition in the models that posses a reentrant semimetal in two dimensions. Note that in three dimensions due to the diffusive states at finite energy that dominate the dynamics this probe is not useful[S6]. We initialize a wavepacket to the origin $|\psi_0\rangle$ and use the KPM to time evolve the state to obtain $|\psi(t)\rangle = e^{-iHt}|\psi_0\rangle$ and from this we compute the spread of the wavepacket as a function of time from $\langle r(t)^2 \rangle \equiv \langle \psi(t)|r^2|\psi(t)\rangle$. From the long time dynamics we extract the dynamical exponent from the scaling $\langle r(t)^2 \rangle \sim t^{2/z}$. We focus on the perfect SOC Hamiltonian from the main text. The KPM expands the time evolution operator in terms of Chebyshev polynomials up to an order $N_C$, which dictates the final time that can be reached in the numerical calculations. Here we focus on a large linear system size of $L = 987$ and use a KPM expansion order $N_C = 2^{12}$, which allows us to time evolve the wavepacket until it spreads out across the sample. As shown in Fig. S10 we find rather unusual wavepacket dynamics which is a signature of a sequence of semimetal-metal-semimetal transitions. As a function of increasing $W$ we find that the speed at which the wavepacket spreads out monotonically decreases until we reach the metallic phase where the dependence on $W$ is rather weak. Then, upon reentering the semimetal phase at larger $W$ the wavepacket spreading speeds back up. This is clearly demonstrated in the dynamical exponent $z$ showing non-monotonic behavior in the inset. Interestingly, our estimate of $z$ is not diffusive, consistent with the marginal nature of 2D. We expect that this wavepacket signature can be used to detect the transition in cold atom experiments.

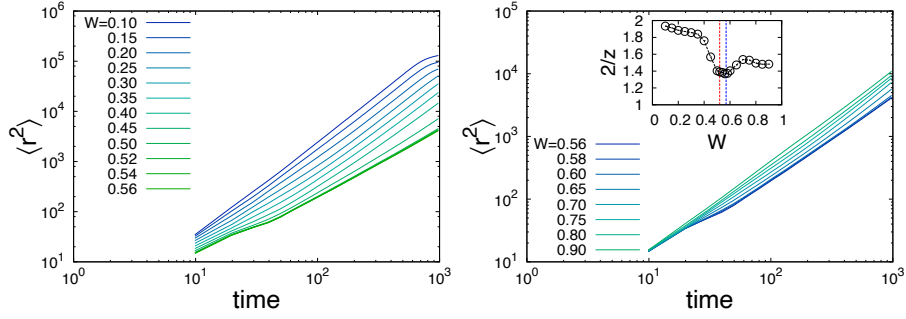

FIG. S10: Wave packet dynamics displaying the spread of the wavepacket $\langle r(t)^2 \rangle$ as a function of time computed from the KPM on a system size of $L = 987$ with a KPM expansion order $N_C = 2^{12}$. (Left) Wavepacket spreading for $W$ in the semimetal and passing into the metallic phase at $W \approx 0.54$. (Right) Similar results but for $W$ passing from the metallic phase to the reentrant semimetal at $W \approx 0.57$. The data clearly shows a non-monotonic wavepacket spreading for increasing quasiperiodic potential strength which gives rise to a non-monotonic behavior in the dynamic exponent $z$ shown in the inset, which we extract via fits to the long time dynamics (the dashed lines mark the entry into the metallic phase).

## B. Wavepacket expansion in the hydrodynamic regime of strong interactions

This section is devoted to the dynamics of a cold atomic cloud in magic-angle emulators. We concentrate on the model of perfect SOC in 2D, and incorporate velocity and wave function renormalization perturbatively, as exposed in Sec. III A. Complementary to the previous section, we here consider strong interactions leading to fast equilibration, justifying an effective hydrodynamic approach.

We describe the dynamics of the system using the Boltzmann equation for the distribution function $f_n$

$$\dot{f}_n + \dot{\boldsymbol{r}}_n \nabla_r f_n + \dot{\boldsymbol{p}} \nabla_p f_n = \mathrm{St}_n[f]. \tag{S37}$$

Here, we introduced the multi index $n = (\boldsymbol{K}, \boldsymbol{p}, \xi)$ to label the states (node $\boldsymbol{K}$, momentum relative to the node $\boldsymbol{p}$, band $\xi = \pm 1$). Semiclassical fermions have energy $\epsilon_n = \xi[v|\boldsymbol{p}| - v_3|\boldsymbol{p}|^3 - v_{C_4}(p_x^4 + p_y^4)/|\boldsymbol{p}|]$ and velocity $\dot{\boldsymbol{r}}_n = \partial \epsilon_n / \partial \boldsymbol{p}$. The momentum satisfies $\dot{\boldsymbol{p}}_n = \boldsymbol{F}(\boldsymbol{r})$ for force $\boldsymbol{F}$. The collision kernel (outgoing states are denoted by a bar) is

$$St_n[f] = - \sum_{m, \bar{n}, \bar{m}} \left\{ W_{nm \to \bar{n}\bar{m}} f_n f_m (1 - f_{\bar{n}})(1 - f_{\bar{m}}) - W_{\bar{n}\bar{m} \to nm} f_{\bar{n}} f_{\bar{m}} (1 - f_n)(1 - f_m) \right\}. \tag{S38}$$

The Boltzmann treatment is valid when either thermal length or Fermi wave length exceed the mean free path. We determine the transition probability in the Fermi golden rule approximation (see e.g. Ref. S8)

$$W_{nm \to \bar{n}\bar{m}} = 2\pi U^2 |\langle n|\bar{n} \rangle|^2 |\langle m|\bar{m} \rangle|^2 (2\pi)^3 \delta(\epsilon_n + \epsilon_m - \epsilon_{\bar{n}} - \epsilon_{\bar{m}}) \delta(\boldsymbol{p}_n + \boldsymbol{p}_m - \boldsymbol{p}_{\bar{n}} - \boldsymbol{p}_{\bar{m}}) \tilde{\delta}_{\boldsymbol{K}_n + \boldsymbol{K}_m, \boldsymbol{K}_{\bar{n}} + \boldsymbol{K}_{\bar{m}}}. \tag{S39}$$

The tilde on the Kronecker delta denotes equality modulo $2\pi$. The collision integral conserves particle number and energy $\sum_n \mathrm{St}_n[f] = 0 = \sum_n \epsilon_n \mathrm{St}_n[f]$. While it does not conserve the total crystalline momentum $\boldsymbol{p}^{\mathrm{tot}} = (\boldsymbol{K} + \boldsymbol{p})$ due to umklapp scattering, the momenta $\boldsymbol{p}$ relative to the nodes are conserved $\sum_n \boldsymbol{p}_n \mathrm{St}_n[f] = 0$. Therefore, number density $\rho_N = \sum_n f_n$ and energy density $\rho_E = \sum_n \epsilon_n f_n$ are conserved and the energy current $\dot{\boldsymbol{j}}_\epsilon = \sum_n \epsilon_n \dot{\boldsymbol{r}}_n f_n$ ($\epsilon_n \dot{\boldsymbol{r}}_n = v^2 \, \boldsymbol{p}_n + \mathcal{O}(v_3, v_{C_4})$) decays slowly due to nonlinearities in the spectrum. We estimate its decay rate as $1/\tau = \Gamma_0 f(\alpha) Z(\alpha)^2 / \bar{v}(\alpha)^2$ where $\Gamma_0 = U_0^2/t$ and $f(\alpha)$ is a dimensionless regular function of $\alpha$ which is proportional to $v_3, v_{C_4}$.

Following a formalism of hydrodynamics in graphene[S9] we introduce smoothly varying Lagrange multipliers (chemical potential, temperature, hydrodynamic velocity) for each of the conserved quantities. The Fermi energy (measured from the Dirac node) is assumed to be small as compared to the mini-bandwidth. In particular the hydrodynamic velocity $\boldsymbol{u}$ enters as $\boldsymbol{j} = \boldsymbol{u}\rho_N$, and for $\mu \ll T$, $\rho_N \sim \mu T/v^2 + \mathcal{O}(u^2/v^2)$, $\rho_E \sim T^3/v^2 + \mathcal{O}(u^2/v^2)$. The hydrodynamic equations reduce to

$$\dot{\rho}_N + \nabla_r(\rho_N \boldsymbol{u}) = 0 \tag{S40a}$$

$$\dot{\rho}_E + \nabla_r \left( \frac{3v^2}{2v^2 + u^2} \boldsymbol{u} \rho_E \right) = \boldsymbol{F} \cdot \boldsymbol{u} \rho_N \tag{S40b}$$

$$\boldsymbol{u} \partial_t P + v^2 \nabla P + \mathcal{H}(\partial_t \boldsymbol{u} + (\boldsymbol{u} \cdot \nabla)\boldsymbol{u}) = \rho_N(v^2 \boldsymbol{F} - \boldsymbol{u}(\boldsymbol{F} \cdot \boldsymbol{u})) - \frac{\mathcal{H}\boldsymbol{u}}{\tau} \tag{S40c}$$

The quantity $\mathcal{H} = \frac{3v^2}{2u^2 + v^2}\rho_E$ is the enthalpy and $P = \frac{v^2 - u^2}{2v^2 - u^2}\rho_E$ is the pressure. The first two formulae are continuity equations, while the last expression is the Euler equation for relativistic fluids. We have neglected all dissipative (viscous) corrections. We incorporate the magic angle effect by using $v = v(\alpha)$ and the force $\boldsymbol{F} = -UZ(\alpha)^2 \nabla \rho_N$ accounts for onsite Hubbard interaction energy in the Hartree approximation.

We numerically solve Eqs. (S40), see Fig. S11 which demonstrates the manifestation of the magic angle effect in various quantities: (i) There is an $\alpha$ dependent crossover from a short ballistic evolution to diffusive motion at time $\tau$. (ii) Near the magic angle, the overdamped dynamics [vanishing $\tau(\alpha)$] prevents the ballistic wave front propagation of energy density (compare c) vs. d)). (iii) Similarly, in prevents the formation of waves in the density profile and (iv) the negative expansion velocity of average position $\langle r^2 \rangle$ at intermediate time. (v) The behavior as a function of $\alpha$ (or $W$) is nonmonotonic, compare Fig. S11 e),f) and Fig. S10.

## C. Discussion of experimental observations

We propose the observation of wave packet dynamics that can be measured in experiment using absorption imaging techniques. We have discussed two complementary limits of wavepacket dynamics: (i) the non-interacting limit, Fig. S10 (ii) an interacting Fermi gas, Fig. S11.

The experimental time evolution is limited, e.g. due to particle loss, to about 50 hopping times. This is sufficient to resolve the onset of the crossover from ballistic to (super-)diffusive motion in the non-interacting case.

In the strong coupling regime, and for the chosen parameters in Fig. S11 ($\rho_E = v(\alpha = 0)/\xi^2$) the energy density at $\alpha = 0$ implies a typical temperature scale of the order $T \sim ta/\xi$ ($a$ is the lattice constant). For $a/\xi \sim 1/4$ this corresponds to $T/T_F \sim 1/4$ which is well in the experimentally realistic regime. The observation of the consequences of the magic angle effect at fixed energy density requires a time evolution at least up to $10\xi/v(\alpha = 0) = 10(\xi/a)/t$. Thus, for typical cloud sizes $\xi \sim 4a$ (i) the typical temperature in Fig. S11 is much smaller than the bandwidth (ii) magic angle effects are well resolvable in present day cold atomic experiments with coherence up to 50 hopping times.

## V. MULTIFRACTAL ANALYSIS AND WAVEFUNCTIONS

In this section, we explain the procedure of extracting the multifractal exponents and the underlying physical interpretations.

## A. Scaling exponent $\tau_M$

We now discuss how we extract $\tau_M(q)$ from inverse participation ratio (IPR). For a given wavefunction in momentum space, the finest grid is $2\pi/L \times 2\pi/L$. We introduce an integer binning factor $B$ which controls the resolution of the momentum space wavefunction. The IPR constructed in this way has been dubbed the momentum space IPR[S6] and given by

$$\mathcal{I}_M(q, L; B) = \sum_{k'} \left| \tilde{\phi}_E(k'; B) \right|^{2q} \sim \left( \frac{Ba}{L} \right)^{\tau_M(q)},$$ (S41)

where $k'$ denotes the wavevector of the binned wavefunction grid and $\tilde{\phi}_E(k'; B)$ is the binned wavefunction. The scaling behavior only holds when $1 \ll B \ll L/a$. To extract $\tau_M(q)$, we choose two consecutive values of binning factors, $B_1$ and $B_2$, and then perform numerical derivative as follows:

$$\tau_M(q; B_1, B_2) = \frac{\ln \mathcal{I}_M(q, L, B_1) - \ln \mathcal{I}_M(q, L, B_2)}{\ln B_1 - \ln B_2}.$$ (S42)

At last, we average $\tau_M(q; B_1, B_2)$ over phases in the quasiperiodic potential. For $L = 144$, we construct $\tau_M$ with 100 realization. We only use 10 realization for $L = 610$. For real space scaling exponents ($\tau_R$), we adopt similar procedures to extract their values. For the case of cTBG, we use 100 realizations of $L = 377$, but to handle binning we truncate the wavefunction at the edge of the BZ (where the wave function is effectively zero) to produce a $376 \times 376$ wave function we can then perform the binning scheme on.

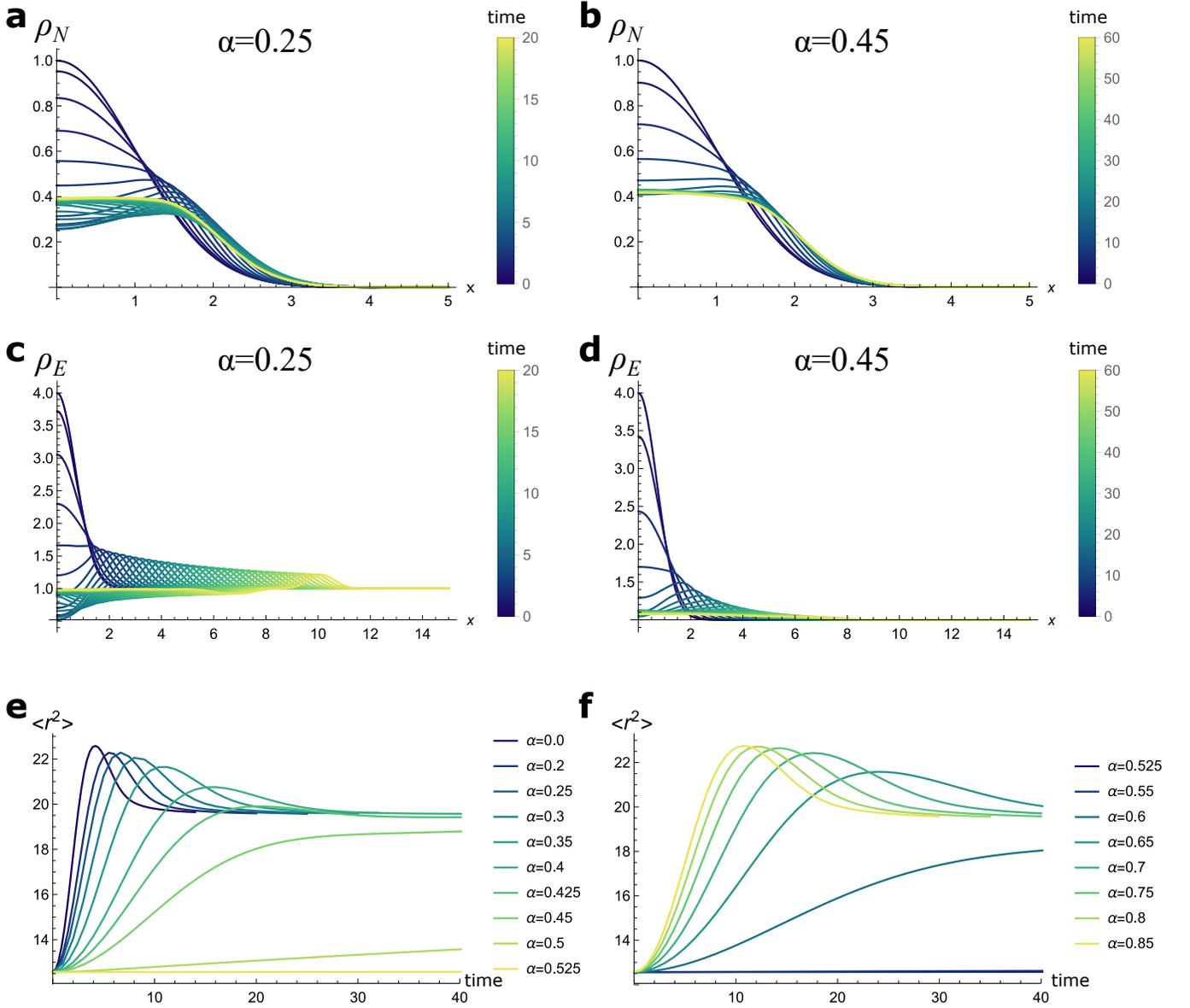

FIG. S11: Simulation of hydrodynamic evolution, Eq. (S40). Initial conditions are $\rho_E(0) = \frac{v(\alpha=0)}{\xi^3}\left(1 + 3e^{-r^2/\xi^2}\right)$, $\rho(\boldsymbol{r}, t=0) = \frac{e^{-r^2/[2\xi^2]}}{\xi^2}$, and $\boldsymbol{u}(0) = 0$. In the plots, all lengths are in units of $\xi$, all times in units of $\xi/v(\alpha=0)$, in these units $U(\alpha=0) = 0.1, 1/\tau(\alpha=0) = 0.03$. Evolution for various times (shown as color) of the particle density a) far from and b) near the magic angle $\alpha_c = 0.536$; c) d) similarly for the evolution of the energy density. e), f): Evolution of $\langle r^2 \rangle = \int d^2x\, r^2 \rho_N(\boldsymbol{r}, t)$, where the value of $\alpha$ is shown as color.

## B. Real space multifractality

As a complementary analysis, we study the multifractal spectrum of the real space wavefunctions at zero energy. We construct the corresponding real space scaling exponent spectrum $\tau_R(q)$ which encodes plane wave, multifractal, and localized wavefunctions.

For $W/t < 0.525$, the real space wavefunction is a plane wave, characterized by the Fourier modes at Dirac points $(k_x, k_y) = (0,0), (0,\pi), (\pi,0)$, and $(\pi,\pi)$. The multifractal spectrum $\tau_R(q)$ is characterized by a straight line $\tau_R(q) = 2(q-1)$. Plane wave states can also be found for $0.55 < W < 1.5$, the inverted semimetallic regime. For $0.525 < W/t < 0.55$, the real space wavefunctions show multifractal behavior characterized by a nonlinear $\tau_R(q)$ spectrum. For $W/t \gg 1.5$, the real space wavefunctions become localized, $\tau_R(q) = 0$ for $q > 0$.

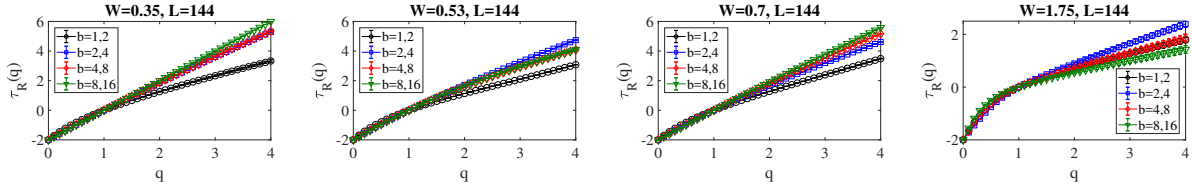

FIG. S12: $\tau_R(q)$ with different values of $W$ with $L = 144$ and 100 realizations. For $W = 0.35$ and $W = 0.7$, the multifractal spectra approach to plane wave, $\tau_R(q) = 2(q-1)$, as increasing $b$. For $W = 0.53$, the spectrum is multifractal described by a nonlinear function. For $W = 1.75$, $\tau_R$ approaches to a localized spectrum, $\tau_R(q) = 0$ for $q > 0$, as increasing $b$. The existing data still suffer from a strong finite size effect and do not clearly show a localized spectrum. We therefore turn to the finite size dependence of the real space wavefunction with $W = 1.75$, which indeed shows localization as the real space IPR is $L$ independent in this regime.

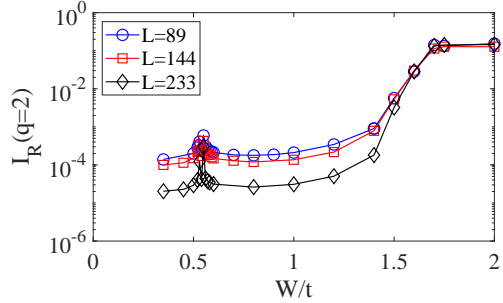

FIG. S13: The inverse participation ratio of real space wavefunctions with different system sizes. Blue circles represent the data of $L = 89$ with 100 realization; red squares represent data of $L = 144$ with 100 realization; black diamonds represent data of $L = 233$ with 50 realization. The collapse of data with different sizes ($W/t \geq 1.75$) indicate localized regime.

We numerically compute the $\tau_R(q)$ spectrum for various values of $W$ in different phases. In Fig. S12, we compute $W/t = 0.35, 0.53, 0.7, 1.75$ for $L = 144$ and average over 100 realizations. $W/t = 0.35$ and $W/t = 0.7$ show plane wave spectrum as increasing the real space binning factor $b$. $W/t = 0.53$ demonstrate a weakly nonlinear multifractal spectrum with a fitting function $\tau_R(q) \approx 2(q-1) - 0.16q(q-1)$ for $|q| < 1$. A much larger system size is needed for quantitatively determining the spectrum which is beyond the scope of this work. For $W/t = 1.75$, the spectrum gradually approaches to localized like behavior as increasing $b$. In our $L = 144$ data, we do not find a clear localized spectrum, due to finite size effects. We therefore turn to the real space IPR data as a function of system size in Fig. S13. This indeed shows strong evidence for localization as the IPR is $L$ independent for $W/t \geq 1.75$.

## C. Interpretation of unfreezing transition

As we have demonstrated in the maintext, the momentum space wavefunctions serve as a proxy for the semimetal-metal transition. The multifractal analysis in momentum space here is distinct from the conventional notion of wavefunctions in real space[S10]. We discuss the interpretation of the multifractal spectrum in depth here.

For $W = 0$, the zero energy wavefunction is composed of the Fourier modes at the Dirac points . The zero energy states are linear combinations of these four plane waves. The probability distributions (integrating over the spin degrees of freedom) of the momentum space wavefunction generically contains four peaks, which we call "ballistic peaks". In the multifractal $\tau_M(q)$ spectrum, these ballistic peaks correspond to a frozen spectrum, $\tau_M(q) = 0$ for $q > 1$.

The frozen feature in the momentum space wavefunction suggests that the ballistic peaks are sharply defined, with the finest localization length $\pi/L$. The $\tau_M(q)$s extracted from the numerics weakly depend on the choice of binning factors $B_1$ and $B_2$. As $W$ increases, satellite peaks with weaker amplitude arise. When $W \approx W_c$, the $\tau_M(q)$s still show freezing behavior for $B = 1, 2$ but become generically non-zero for larger binning factors. Such features suggests that the distance between a ballistic peak and the nearest satellite peaks is around $2 * 2\pi/L$ to $4 * 2\pi/L$. The ballistic peaks start to hybridize with the satellite peaks when $W \geq W_c$. This corresponds to an unfreezing transition in momentum space, which describes a zero-measure set to an extensive set of Fourier modes. The $W_c$ determined here

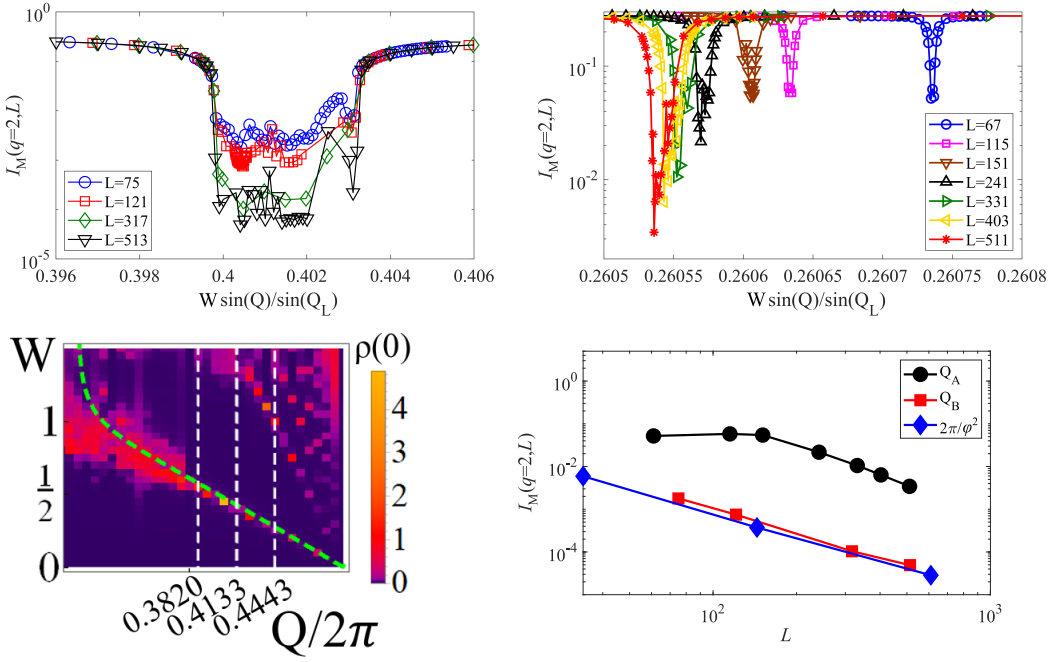

FIG. S14: Comparisons of IPRs in momentum space that show scaling with system size indicative of delocalization in momentum space, (Top left) $Q = Q_B = 2\pi(51 - \sqrt{5})/181 \approx 2\pi \times 0.4133$ (Top right) $Q = Q_A = 2\pi(1/2 - 1/\varphi^6) \approx 2\pi \times 0.4443$. (Bottom left) The phase diagram, as in Fig 1 **b** of the main text, with cuts along $Q = 2\pi/\varphi^2 \approx 2\pi \times 0.3820$, $Q = Q_B = 2\pi \times 0.4133$, $Q = Q_A = 2\pi \times 0.4443$. (Bottom right) Summary of system size dependence of the minimal IPR along these three cuts, demonstrating the existence of a crossover scale between delocalized and localized regimes for $Q = Q_A$.

coincides well with the semimetal-metal transition extracted from the density of states calculations.

For larger values of $W$, an inverse semimetal transition, metal-to-semimetal, takes place. The same multifractal analysis in momentum space also applies.

### D. Minimal system size for momentum space delocalization

In this section we discuss the minimal system size required to observe the universal magic angle effect, i.e. momentum space delocalization at the magic angle.

At small twist angle, the separation of length scales between original lattice spacing and moiré supercell becomes large and suppresses the size of the metallic sliver in the phase diagram, Fig. 1**b**, **c** of the main text. This renders the observation in numerical (i.e. finite size) simulations very difficult. At smallest twist angles, $\theta \ll 10°$ for TBG and $(\pi - Q) \ll 1$ for 2D SOC, wavefunction delocalization is not observable at small system sizes.

However, for the specific case of the 2D SOC model and a given $Q \approx \pi$, we here demonstrate the appearance of a crossover length scale, beyond which the magic angle effect becomes observable. In Fig. S14 we present data for $Q = 2\pi(1/2 - 1/\varphi^6) \approx 2\pi \times 0.4443$, for which wave function delocalization (i.e. system size dependent IPR) occurs for $L \gtrsim 160$ and we compare this to other $Q$ further from $\pi$. Note that for $Q \approx 2\pi \times 0.4443$ the position of the magic angle non-trivially depends on the finite size (i.e. on the concrete numerical rational approximant).

We conclude the discussion of the origin of this critical system size with a physical interpretation. Within our finite size numerics, the incommensurate perturbation repeats itself on the length scale of the entire system. Within our picture, the commensurate structure provides a cut-off for the single particle transition that rounds out the non-analyticity and the wavefunctions do not delocalize in momentum space. Based on this and our numerical observation, we conjecture that for any irrational $Q$ and sufficiently large system size, wave function delocalization may be observed.

## VI. WANNIER STATES AND BUILDING HUBBARD MODELS

In order to analytically build the Hubbard models analyzed in the main text, we use the Hamiltonian as perviously defined

$$
\begin{aligned}
\hat{H} &= \hat{T} + \hat{V} + \hat{U}, \\
\hat{T} &= \sum_{\mathbf{r},\mu} [\frac{it}{2} c_{\mathbf{r}}^{\dagger} \sigma_{\mu} c_{\mathbf{r}+\hat{\mu}} + \text{h.c.}], \\
\hat{V} &= W \sum_{\mathbf{r},\mu} \cos(Q_n r_\mu + \phi_\mu) c_{\mathbf{r}}^{\dagger} c_{\mathbf{r}}, \\
\hat{U} &= U \sum_{\mathbf{r}} c_{\mathbf{r},\uparrow}^{\dagger} c_{\mathbf{r},\downarrow}^{\dagger} c_{\mathbf{r},\downarrow} c_{\mathbf{r},\uparrow}.
\end{aligned}
\tag{S43}
$$

We take $Q_n = 2\pi F_{n-2}/F_n$ and the system size to be $L = mF_n$ such that $F_n$ defines a supercell of size $\ell = F_n$. The model has a particle-hole symmetry when $\phi_\mu = \pi/2$, which we will concentrate on in order to isolate bands near $E = 0$.

The first task is to isolate bands. With $L = m\ell$, our Brillioun Zone will have $m^2$ sampled points. Due to the imposed discrete translational symmetry, one can write the single particle Hamiltonian $\hat{H}_{\text{sp}} = \hat{T} + \hat{V}$ as

$$
\hat{H}_{\text{sp}} = \bigoplus_{n_x, n_y = 0}^{m-1} \hat{H}_{\mathbf{n}},
\tag{S44}
$$

for $\mathbf{n} = (n_x, n_y)$ and, using Bloch's theorem, the individual Hamiltonians are (in first quantized notation)

$$
\hat{H}_{\mathbf{n}} = \sum_{\mathbf{r},\mu} \left[ \frac{it}{2} e^{-i2\pi n_\mu / L} |\mathbf{n}, \mathbf{r}\rangle \langle \mathbf{n}, \mathbf{r} + \hat{\mu}| \otimes \sigma_\mu + \text{h.c.} \right] + W \sum_{\mathbf{r},\mu} \cos(Q_n r_\mu + \phi_\mu) |\mathbf{n}, \mathbf{r}\rangle \langle \mathbf{n}, \mathbf{r}|,
\tag{S45}
$$

on a system size of size $\ell$. In the thermodynamic limit $m \to \infty$, and we can plot dispersions to see if a gap has opened. Once diagonalized, we have a set of states $\{E_{\mathbf{n},j}\}$, and looking near $E = 0$, we can form projectors onto energy states within a miniband

$$
\hat{P} = \sum_{E_{\mathbf{n},j} \in \text{Miniband}} |\mathbf{n}, j\rangle \langle \mathbf{n}, j|.
\tag{S46}
$$

This projector should have an integer multiple of $m^2$ states within it, and each $(n_x, n_y)$ pair should contribute the same number of states. This is a check to determine if we have a "good band."

Further, we can get an idea of where the Wannier centers should be by looking at the integrated local density of states

$$
\rho_{\text{Band}}(\mathbf{r}) = \langle \mathbf{r} | \hat{P} | \mathbf{r} \rangle.
\tag{S47}
$$

Considering systems with periodic boundary conditions, we consider the position operators $\{\cos(2\pi \hat{x}/L), \sin(2\pi \hat{x}/L), \cos(2\pi \hat{y}/L), \sin(2\pi \hat{y}/L)\}$, and we construct the projected operators

$$
\{\hat{P} \cos(2\pi \hat{x}/L) \hat{P}, \hat{P} \sin(2\pi \hat{x}/L) \hat{P}, \hat{P} \cos(2\pi \hat{y}/L) \hat{P}, \hat{P} \sin(2\pi \hat{y}/L) \hat{P}\}.
\tag{S48}
$$

These operators no longer commute with one another and cannot be simultaneously diagonalized. The solution is *approximate joint diagonalization* (AJD) achieved by minimizing a cost function: for a set of operators $\mathcal{O}$, we need to find a unitary matrix $U$ such that

$$
\sigma(U) = \sum_{A \in \mathcal{O}} \sum_{i \neq j} |(U^{\dagger} A U)_{ij}|^2
\tag{S49}
$$

is minimized. An efficient algorithm for this was developed for signal processing[S11] and enumerated for maximally localized Wannier states in Ref. S12. Our algorithm to achieve this, written in Julia, can be found in Ref. S13.

The columns of the resulting unitary matrix $U$ are the maximally localized Wannier states. The resulting single particle part of tight-binding Hamiltonian is found numerically to be given approximately by

$$
\hat{H}_{\text{Wannier}} = U^{\dagger} \hat{P} \hat{H}_{\text{sp}} \hat{P} U \approx \sum_{\mathbf{R},\mu} \left[ \frac{it_{\text{eff}}}{2} |W_{\mathbf{R},\tau}\rangle \langle W_{\mathbf{R}+\hat{\mu},\tau'}| [\sigma_\mu]_{\tau,\tau'} + \text{h.c.} \right],
\tag{S50}
$$

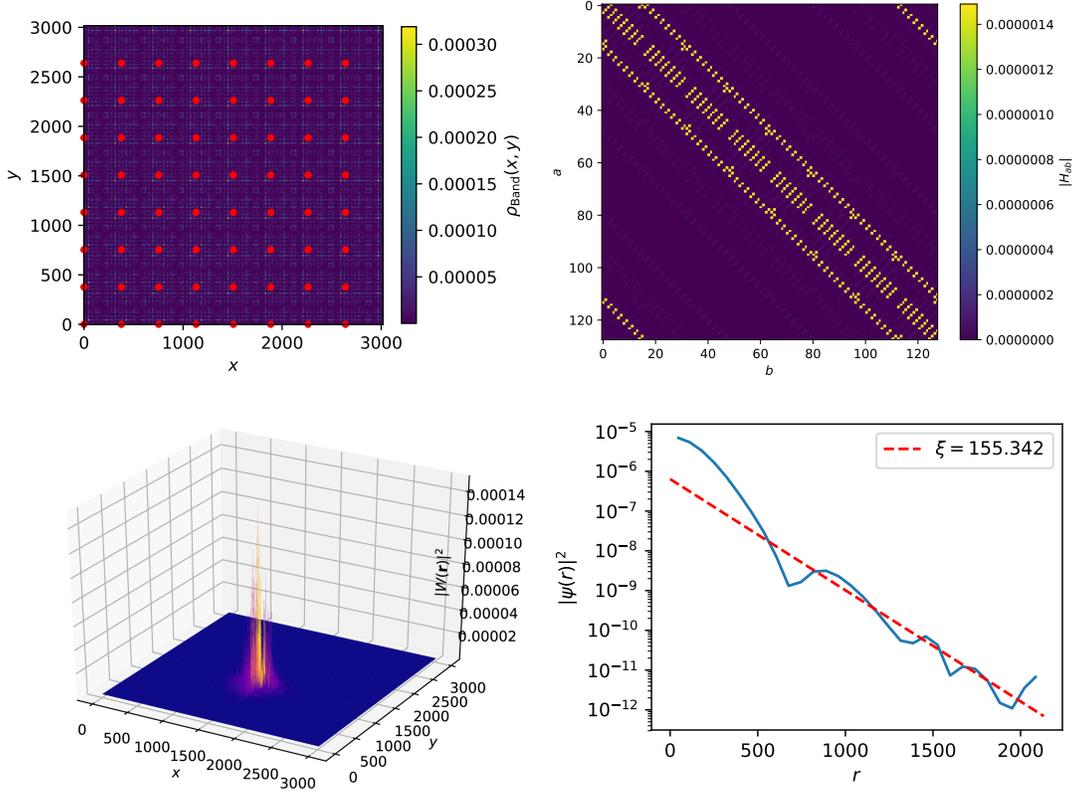

FIG. S15: (Upper left) The integrated density of states $\rho_{\text{Band}}(x, y)$ along with Wannier centers (red dots) for $\ell = 377$ and $m = 8$ and $W = 0.52445t$. (Upper right) The magnitude of the elements of the Wannier Hamiltonian $H_{\text{Wannier}}$. (Lower left) 3D representation of a single Wannier function $|W(\mathbf{r})|^2$ and (Lower right) The exponential localization of the Wannier function $|W(\mathbf{r})|^2 \sim e^{-r/\xi}$.

for Wannier states $|W_{\mathbf{R}, \tau}\rangle$ labeled by the emergent lattice positions $\mathbf{R}$ and internal index $\tau = 1, 2$. The interaction term also gets renormalized (again, returning to second quantized notation)

$$\hat{H}_{\text{Wannier-int}} = U_{\text{eff}} \sum_{\mathbf{R}} d^{\dagger}_{\mathbf{R}, 1} d^{\dagger}_{\mathbf{R}, 2} d_{\mathbf{R}, 2} d_{\mathbf{R}, 1} \tag{S51}$$

where $d_{\mathbf{R}, \tau}$ annihilates the states $|W_{\mathbf{R}, \tau}\rangle$, and

$$U_{\text{eff}} = U \sum_{\mathbf{r}} \{|W^{\uparrow}_{\mathbf{R}, 1}(\mathbf{r})|^2 |W^{\downarrow}_{\mathbf{R}, 2}(\mathbf{r})|^2 + |W^{\uparrow}_{\mathbf{R}, 2}(\mathbf{r})|^2 |W^{\downarrow}_{\mathbf{R}, 1}(\mathbf{r})|^2 - 2\Re[W^{\uparrow}_{\mathbf{R}, 1}(\mathbf{r})^* W^{\downarrow}_{\mathbf{R}, 2}(\mathbf{r})^* W^{\downarrow}_{\mathbf{R}, 1}(\mathbf{r}) W^{\uparrow}_{\mathbf{R}, 2}(\mathbf{r})]\}, \tag{S52}$$

where $\langle \mathbf{r}, \sigma | W_{\mathbf{R}, \tau} \rangle = W^{\sigma}_{\mathbf{R}, \tau}(\mathbf{r})$ and $U_{\text{eff}}$ is found to be (numerically) independent of $\mathbf{R}$.

To perform the calculations, we look close to the transition $W_c$ calculated previously and find where the smallest gap opens in the dispersion. In the range where that gap opens and closes, we perform Lanczos to find the Bloch states and then we perform AJD to find the Wannier states. After making sure the states are "good" states (exponentially localized and make the appropriate emergent lattice), we can find the Hamiltonian as described above.

### A. Computed Wannier functions

As described in Sec. III C, we can probe minibands by using successive pairs of odd Fibonaccis (even Fibonaccis gap out the Dirac nodes, not allowing us to make the effective Hubbard model described above). We perform the calculation for $\ell = F_n = 13, 21$ (second miniband), $\ell = F_n = 55, 89$ (third miniband), and $\ell = 233, 377$ (fourth miniband). In the main text, we present data for $\ell = 13$, and here we show data for the point of highest $U_{\text{eff}}/t_{\text{eff}} \approx 4115 U/t$.

After tracking the gap opening and closing when $\ell = 377$, we construct the Wannier states (at $W = 0.52445t$ and $m = 8$), and the results are in Fig. S15. We see Wannier centers on top of the $\rho_{\text{Band}}(\mathbf{r})$ in Fig. S15 upper-left, along

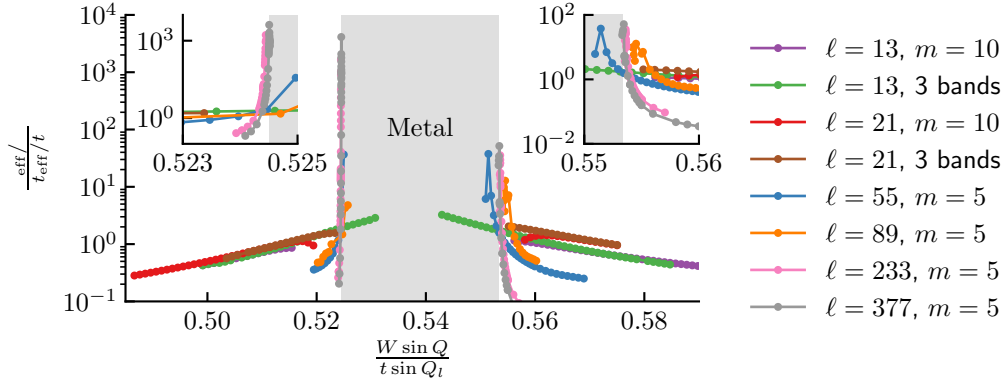

FIG. S16: The interaction of the center band approaching the transition. Each successive pair of odd $\ell$ represent the opening of a new miniband. The data labeled "3 bands" comes from the mulitorbital models, following the Wannier states continuously connected to the single-orbital model. Note that $Q = 2\pi/\varphi^2$ for golden ratio $\varphi$ and $Q_\ell = 2\pi F_{n-2}/\ell$ with $\ell = F_n$.

with a visual representation of the resulting $H_{\mathrm{Wannier}}$ on the upper-right (clearly showing a banded Hamiltonian). The Wannier state is visually seen in Fig. S15 lower-left and after we bin the data, we find the states are exponentially localized in Fig. S15 lower-right.

Doing this multiple times leads to an estimate of the effective interactions on either side of the metal transition in Fig. S16. We should note that for $\ell = 13, 21$, part of the data is pulled from a multi-band Hubbard model by tracking $U_{\mathrm{eff}}/t_{\mathrm{eff}}$ on the Wannier functions with maximal overlap prior to the multi-band Hubbard model. Also, the Wannier states for $\ell = 89$ could not fully be fully converged, so the data for $U_{\mathrm{eff}}/t_{\mathrm{eff}}$ is appropriately noisy (and an underestimate).

## B. Finite energy bands and multi-orbital Hubbard model

It is interesting to note that at finite energy, flat bands develop and in some cases intersect the band near $E = 0$ which is the focus of this text. As a clear example, note for $\ell = 13$, see that a flat, Dirac bands appear around the region labeled 1 in Fig. S17(upper left) and intersect the band in the region labeled 3. These flat bands have greatly increased $U_{\mathrm{eff}}/t_{\mathrm{eff}}$ away from the transition. When the multiorbital Hubbard model appears, the dispersion changes drastically, see Fig. S17(upper right) and has Wannier centers appearing along diagonals within a supercell as seen in Fig. S17(middle left). With the computed data, an entire translationally invariant Hubbard model can be constructed and the result leads to hoppings as indicated in Fig. S17(middle right), from the banded Hamiltonian Fig. S17(lower left). Lastly, the multiorbital Hubbard $U$ is a matrix in this case and is visually represented in Fig. S17(lower right).

---

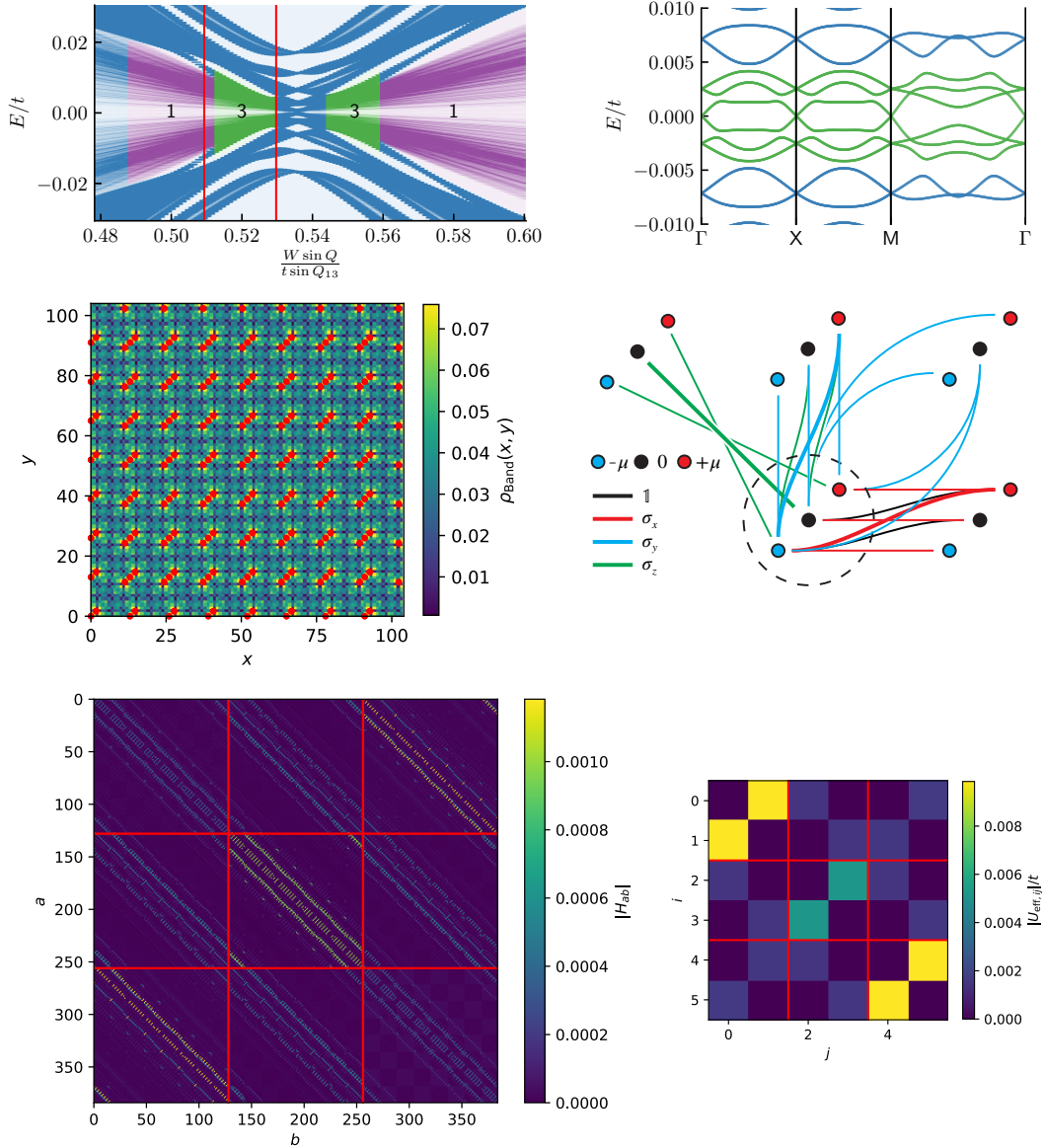

FIG. S17: (Upper left) For $\ell = 13$, we see a region (labeled 1) where one Dirac band exists around $E = 0$ (the figures in the main text are constructed here). There is also an extremely flat band at finite energy that intersects the band in the region labeled 3. (This data is taken for $m \to \infty$ by random sampling the Brillouin zone and $\phi_\mu = \pi/2$.) (Upper right) The dispersion in region 3 at $W = 0.52t$, the second red line in the upper left figure. (Middle left) The integrated density of states $\rho_{\mathrm{Band}}(x, y)$ for the three band model. We can see a line of three states tiled in an $8 \times 8$ grid. (Middle right) A visual representation of the single-particle part of the multi-orbital Hubbard model. The Wannier states are separated by their average energy $-\mu$, $0$, $+\mu$, and they have hopping matrix elements to neighbors and next-to-nearest neighbors on the diagonal. Each Wannier site is double degenerate and the hopping matrix elements are Pauli matrices as color coded. (Bottom left) The Wannier Hamiltonian separated into the 3 orbits (the red lines); we see that it is banded and dominated by the hoppings shown in the middle-right figure. (Bottom right) The Hubbard $U$ matrix for the multi-orbital model. The highest values of the Hubbard $U$ matrix comes from states continuously connected to flat bands in the upper-left figure that intersect the middle band in region 3.


\end{document}